\documentclass[a4paper,11pt,preprint,3p]{elsarticle}
\pdfoutput=1   
\usepackage{tikz}
\usetikzlibrary{tikzmark}
\usetikzlibrary{shapes,arrows}
\usepackage{mathtools,bbm}
\usepackage{amsfonts,amsmath}
\usepackage[percent]{overpic}
\usepackage{cancel,url}
\usepackage{appendix}
\usepackage{color,xcolor}
\usepackage{tabularx, booktabs}
\definecolor{darkred}{rgb}{0.4,0.0,0.0}
\definecolor{darkgreen}{rgb}{0.0,0.4,0.0}
\definecolor{darkblue}{rgb}{0.0,0.0,0.4}
\usepackage[bookmarks,linktocpage,colorlinks,
    linkcolor = darkred,
    urlcolor  = darkblue,
    citecolor = darkgreen]{hyperref}
\numberwithin{equation}{section}
\newcolumntype{Y}{>{\centering\arraybackslash}X} 
\newcommand{\ben}{\begin{enumerate}}
\newcommand{\een}{\end{enumerate}}
\newcommand{\bit}{\begin{itemize}}
\newcommand{\eit}{\end{itemize}}
\newcommand{\beq}{\begin{equation}}
\newcommand{\eeq}{\end{equation}}
\newcommand{\bsa}{\begin{subequations}\begin{eqnarray}}
\newcommand{\esa}{\end{eqnarray}\end{subequations}}
\newcommand{\bea}{\begin{eqnarray}}
\newcommand{\eea}{\end{eqnarray}}
\newcommand{\bean}{\begin{eqnarray*}}
\newcommand{\ean}{\end{eqnarray*}}
\newcommand{\nn}{\nonumber \\}
\newcommand{\non}{\nonumber}
\newcommand{\tr}{\mbox{Tr}}
\newcommand{\fig}[1]{Fig.~\ref{#1}}
\newcommand{\eq}[1]{Eq.~(\ref{#1})}
\newcommand{\tab}[1]{Table~\ref{#1}}
\newcommand{\sect}[1]{Sect.~\ref{#1}}

\begin{document}

\title{
\begin{flushright}
\small{
WUB/19-04\\}
\vskip 0.7cm
\end{flushright}
Constrained Hybrid Monte Carlo algorithms for gauge-Higgs models}

\author{Michael G\"unther}
\ead{guenther@math.uni-wuppertal.de}

\cortext[cor1]{Corresponding author}
\author{Roman H\"ollwieser\corref{cor1}}
\ead{hoellwieser@uni-wuppertal.de}

\author{Francesco Knechtli\corref{}}
\ead{knechtli@physik.uni-wuppertal.de}

\address{Department of Mathematics and Comuter Science,\\ 
Department of Physics, Fakult\"at f\"ur Mathematik und Naturwissenschaften,\\ 
Bergische Universit\"at Wuppertal, Gau{\ss}stra{\ss}e 20, 42119 Wuppertal, Germany}

\begin{abstract}
   We develop Hybrid Monte Carlo (HMC) algorithms for constrained Hamiltonian systems of gauge-Higgs models and introduce a new observable for the constraint effective Higgs potential. 
  We use an extension of the so-called Rattle algorithm to general Hamiltonians for constrained systems, which we adapt to the 4D Abelian-Higgs model and the 5D SU(2) gauge theory on the torus and on the orbifold. The derivative of the potential is measured via the expectation value of the Lagrange multiplier for the constraint condition and allows a much more precise determination of the effective potential than conventional histogram methods. With the new method, we can access the potential over the full domain of the Higgs variable, while the histogram method is restricted to a short region around the expectation value of the Higgs field in unconstrained simulations, and the statistical precision does not deteriorate when the volume is increased. 
 We further verify our results by comparing to the one-loop Higgs potential of the 4D Abelian-Higgs model in unitary gauge and find good agreement. To our knowledge, this is the first time this problem has been addressed for theories with gauge fields. The algorithm can also be used in four dimensions to study finite temperature and density transitions via effective Polyakov loop actions.
\end{abstract}
		

\maketitle

\newpage

\tableofcontents

\newpage

\section{INTRODUCTION}

The Brout-Englert-Higgs (BEH) mechanism \cite{Englert:1964et,Higgs:1964ia} explains the generation of the mass of gauge bosons in gauge theories coupled to
  a scalar field called the Higgs field. The Standard Model (SM) of particle physics relies on this mechanism. In 2012 a scalar particle of mass around $125$ GeV was
  discovered at the LHC  accelerator at CERN \cite{ATLAS:2012gk,CMS:2012gu} rendering the SM complete. The masses of the gauge bosons arise by
  Spontaneous Symmetry Breaking (SSB) triggered by the Higgs potential. The origin of the Higgs potential is, as of yet, unknown. Moreover, the mass of the Higgs particle
  has a quadratic sensitivity to a ultra-violet cut-off, the so-called hierarchy problem. These problems suggest that a more fundamental process is behind the
  Higgs mechanism.

An elegant solution is provided by Gauge-Higgs Unification (GHU) models \cite{Manton:1979kb,Fairlie:1979at,Hosotani:1983vn} and relies on the existence
  of extra dimensions. In these models the Higgs field is identified with (some of) the extra-dimensional components of the gauge field. The gauge symmetry of the higher
  dimensional theory protects the Higgs mass from corrections which are quadratic in the cut-off. Moreover, a Higgs potential is generated by loop effects and can
  give a mass to the gauge bosons in the regular four dimensions.
A particular GHU model in terms of a five-dimensional (5D) SU(2) gauge theory where the extra dimension is compactified on an $S^{1}/\mathbb{Z}_{2}$ orbifold was formulated in \cite{Irges:2004gy,Knechtli:2005dw} in the context of lattice field theory. At the fixed points of the orbifold, the gauge group is explicitly broken down to U(1). The theory exhibits SSB \cite{Irges:2006zf,Irges:2006hg} 
in accordance with Elitzur's theorem \cite{Elitzur:1979uv}, via the spontaneous breaking of the so-called stick symmetry \cite{Ishiyama:2009bk,Irges:2013rya}, giving rise to the BEH mechanism. This observation was confirmed in \cite{Irges:2012ih,Irges:2012mp} via semi-analytic mean-field calculations.

\begin{figure}[h]
\centering
\includegraphics[width=0.393\linewidth]{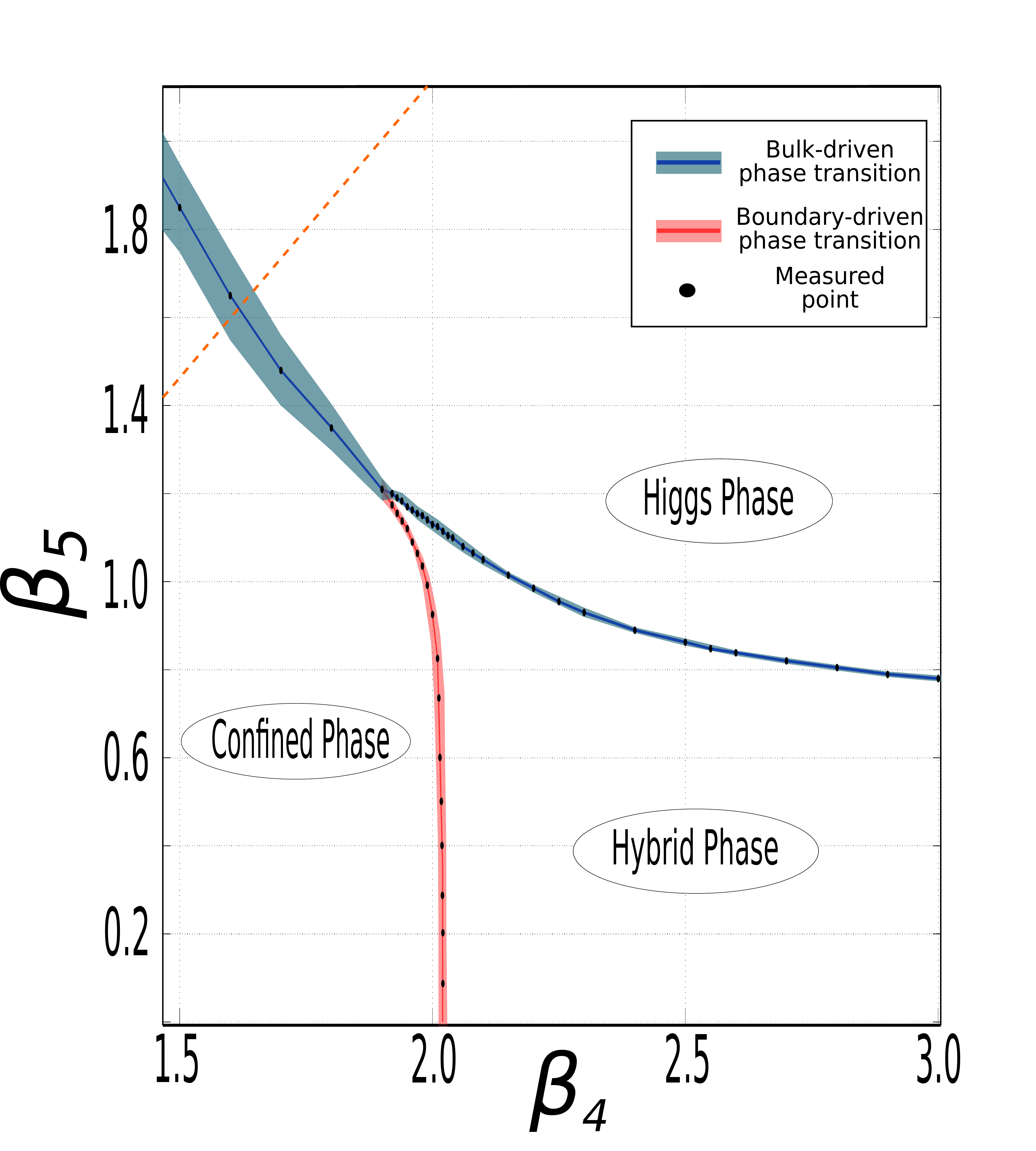}
\includegraphics[width=0.6\linewidth]{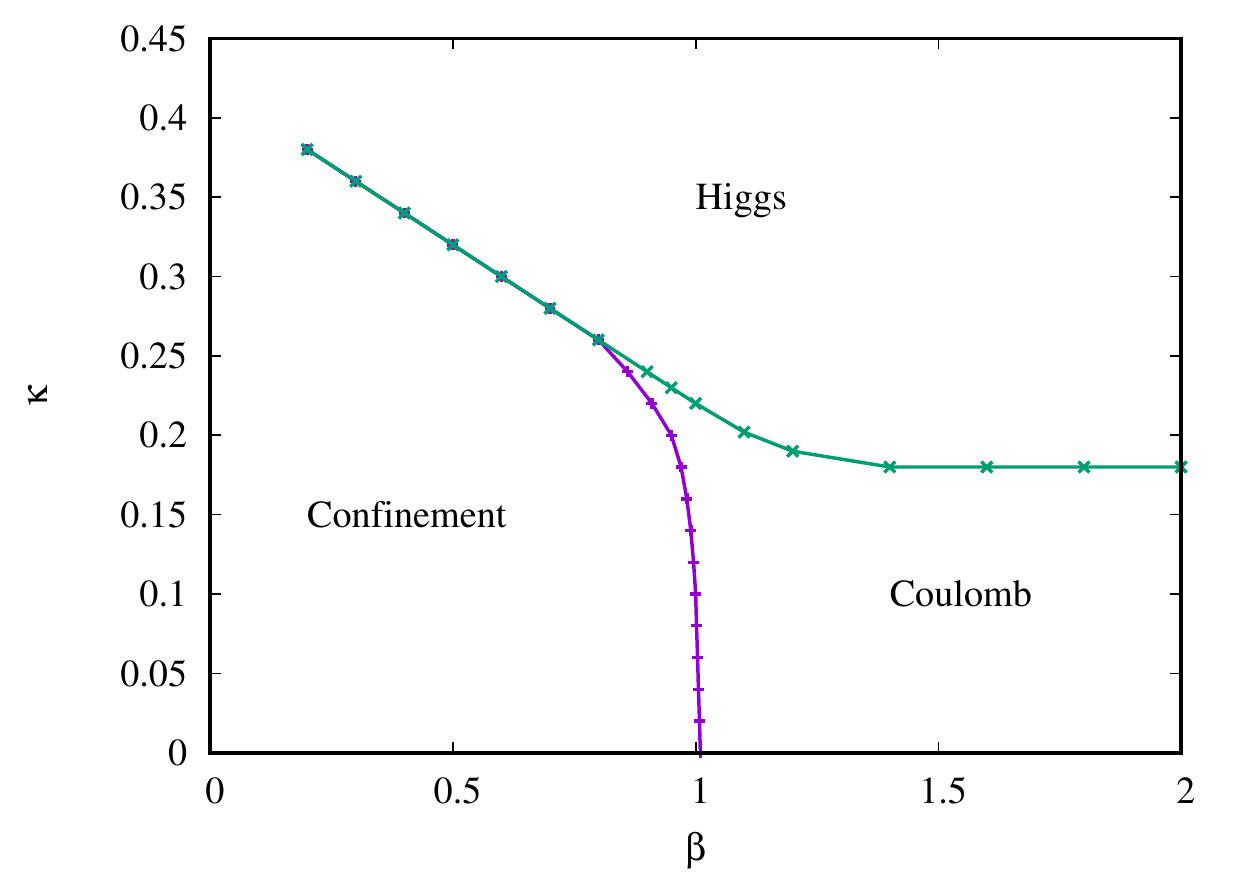}
\caption{\footnotesize (left) From \cite{Alberti:2015pha}. The phase diagram for $N_5=4$ in the region of the Higgs-hybrid phase transition of the 5D orbifold gauge theory (see \sect{sec:chmcorb}). The points show the location of a first-order phase transition. The red and blue lines represent the width of the corresponding hystereses, while the dashed orange line represents $\gamma=1$ ($\beta_4=\beta_5$). (right) The phase diagram of the Abelian-Higgs model  (see \sect{sec:chmcabel}) with $\lambda=1$ showing a similar phase structure. The green and purple curves correspond to different order parameters.}
  \label{fig:pds}
\end{figure}

The system has been found to exhibit three phases, see \fig{fig:pds} (left), 
separated by first order phase transition lines which are characterized by the expectation value of the Polyakov loop in the extra dimension: in the confined (de-confined) phase the Polyakov loop exhibit zero (non-zero) expectation value in every direction. In this context, the de-confined phase is labelled Higgs phase, because it is where the Higgs potential develops SSB, giving rise to non-zero gauge boson masses. The third phase, which is characteristic only of the orbifold geometry, shows confined dynamics in the orbifold's bulk, and de-confined dynamics on its boundaries; it is, therefore, called hybrid phase. These results, which are favorably pointing towards the suitability of this theory for describing the electro-weak sector of the Standard Model, are reported in \cite{Alberti:2015pha,Alberti:2016wff}. 

The phase structure is similar to the one of the four-dimensional (4D) Abelian-Higgs model, shown in \fig{fig:pds} (right). Moreover, on the orbifold boundaries of the 5D GHU model one observes dimensional reduction from five to four dimensions \cite{Alberti:2015pha}, which suggests that there is a localization mechanism for the gauge field.
In order to corroborate the relation of the 5D GHU model with torus and orbifold boundary conditions to the dimensionally reduced theory,
  the 4D adjoint resp. Abelian-Higgs model, we want to compute the effective potentials in the various cases.
 
 The goal of this work is to develop an algorithm for measuring the so-called constraint effective potential in lattice simulations of gauge-Higgs models.
  The constraint effective potential corresponds in the infinite volume limit to the conventional effective potential~\cite{Fukuda:1974ey,ORaifeartaigh:1986axd}.
  A way to measure the constraint effective potential is presented in Ref.~\cite{Kuti:1987bs}. It is based on the Hybrid Monte Carlo (HMC) algorithm~\cite{Duane:1987de}
  for constrained Hamiltonians. The latter include constraint conditions on the Higgs fields which lead to the constrained equations of motion.
  This is discussed in \sect{sec:chmc} where we derive a formula to measure the derivative of the constraint effective potential in terms of the expectation value of the
  Lagrange multiplier for the constraint.
  Then we focus on the implementation of the constrained HMC for the 4D Abelian-Higgs model \sect{sec:chmcabel} and the 5D SU(2) gauge theory with torus \sect{sec:chmctor} and orbifold \sect{sec:chmcorb} boundary conditions, and present constraint effective potentials for all cases.
  In \sect{sec:unit} we compare the constraint effective potential of the 4D Abelian-Higgs model in unitary gauge to the one-loop (continuum) effective potential for this model~\cite{Irges:2017ztc}. 
   In the conclusions \sect{sec:chmcconcl} we give an outlook to the application of our constrained algorithms to measure effective potentials
    in gauge-Higgs models and also in other theories, {\it e.g.}, in finite temperature Quantum Chromodynamics (QCD).

\section{THE CONSTRAINT EFFECTIVE POTENTIAL}\label{sec:chmc}

 The exact effective potential is the infinite volume limit of the so-called constraint potential $U_\Omega(\Phi)\overset{\Omega\rightarrow\infty}{\rightarrow}U_{{\rm eff}}(\Phi)$~\cite{Fukuda:1974ey,ORaifeartaigh:1986axd}.
The latter can be calculated non-perturbatively, via simulating the constrained path integral,
\beq
e^{-\Omega U_\Omega(\Phi)}=\int\mathcal{D}\phi\delta(\frac{1}{\Omega}\sum_{n_\mu}\mathcal H(n_\mu)-\Phi) e^{-S[\phi]}\label{eq:dUint}
\eeq
where $n_\mu$ ($\mu=0,1,2,3$) are the integer coordinates of the points on a lattice with volume $\Omega$ and the average of the Higgs field $\mathcal H(n_\mu)$, constructed from the field variables $\phi(n_\mu)$, takes a fixed value $\Phi$. 
This was first shown in the pure Higgs theory by Kuti and Shen \cite{Kuti:1987bs}, who suggested to measure the derivative of the constraint effective potential $U_\Omega$ with respect to the constraint field $\Phi$ during the constrained simulations. This method requires a separate simulation for every value of $\Phi$, but the effective potential can be determined with greater accuracy than fitting a distribution $P(\Phi)$ from unconstrained simulations. In order to derive $U_\Omega(\Phi)$ we introduce the constrained Hamiltonian
\bea
\tilde H[\phi,\pi]&=&H[\phi,\pi]+\lambda^{(1)}\biggr(\frac{1}{\Omega}\sum_{n_\mu}\mathcal{H}(n_\mu)-\Phi\biggr),\quad H[\phi,\pi]=S[\phi]+\frac{1}{2}\sum_{n_\mu}\pi^2(n_\mu)
\label{eq:cham}
\eea
with fictitious momentum variables $\pi(n_\mu)$, including the Lagrange multiplier $\lambda^{(1)}$, to be determined in such a way that it ensures the constraint condition, which demands that the Higgs field $\mathcal H(n_\mu)$ fluctuates around a fixed average value $\Phi$. Using the constrained Hamiltonian (\ref{eq:cham}) we rewrite the constrained path integral (\ref{eq:dUint}) as
\bea
e^{-\Omega U_\Omega(\Phi)}&=&\int\mathcal{D}\phi\delta(\frac{1}{\Omega}\sum_{n_\mu}\mathcal H(n_\mu)-\Phi) e^{-S[\phi]}=\int\mathcal{D}\phi\mathcal{D}\pi e^{-\tilde H[\phi,\pi]}\label{eq:eUeff}
\eea
The derivative of \eq{eq:eUeff} with respect to the constrained variable $\Phi$ yields
\bea
-\Omega U_\Omega'e^{-\Omega U_\Omega}&=&-\int\mathcal{D}\phi\mathcal{D}\pi \tilde H'e^{-\tilde H}=\int\mathcal{D}\phi\mathcal{D}\pi\lambda^{(1)} e^{-\tilde H}\label{eq:dUeff}\\
\Rightarrow U_\Omega'(\Phi)&=&-\dfrac{1}{\Omega}\dfrac{\int\mathcal{D}\phi\mathcal{D}\pi\lambda^{(1)} e^{-\tilde H}}{e^{-\Omega U_\Omega}}=-\dfrac{1}{\Omega}\big\langle\lambda^{(1)}\big\rangle_\Phi\equiv U_{\Omega,\rm{cnst.}}'\label{eq:numAH}
\eea
the derivative of the constraint effective potential $U_\Omega'(\Phi)\equiv U_{\Omega,\rm{cnst.}}'$ given by the expectation value of the first Lagrange multiplier during simulations at fixed $\Phi$ ($\langle\ldots\rangle_\Phi$).

The simulations are performed using Hybrid Monte Carlo methods~\cite{Duane:1987de} implementing constrained equations of motion (cEOMs) of the form
\bea
\dot\phi(n_\mu)&=&\dfrac{\partial\tilde H}{\partial\pi(n_\mu)}=\pi(n_\mu)\quad\mbox{and}\quad\dot\pi(n_\mu)=-\dfrac{\partial\tilde H}{\partial\phi(n_\mu)}=-\dfrac{\partial S}{\partial\phi(n_\mu)}-\dfrac{\lambda^{(1)}}{\Omega}\dfrac{\partial\mathcal H}{\partial\phi(n_\mu)},
\eea
including a term incorporating the Lagrange multiplier, which has to be evaluated first, before solving the cEOMs. This is done by demanding that the first derivative of the constraint condition with respect to molecular dynamics time, the so-called hidden constraint, vanishes as the constraint is a conserved quantity. 
 For a non-composite Higgs field $\mathcal H(n_\mu)$ we have a constraint condition that is linear in the underlying fields and the hidden constraint only depends on the momenta $\pi(n_\mu)$. In this case we can apply standard leap-frog algorithms, which preserve linear constraints exactly, as all Runge-Kutta schemes. If the constraint is applied to composite fields however, {\it e.g.}, $\mathcal H(n_\mu)=\phi^\dagger(n_\mu)\phi(n_\mu)$ as in the Abelian-Higgs model, we get additional conditions of the form $\sum_{n_\mu}\dot\phi(n_\mu)\phi(n_\mu)=\sum_{n_\mu}\pi(n_\mu)\phi(n_\mu)$, depending on $\pi(n_\mu)$ and $\phi(n_\mu)$. In standard leap-frog algorithms, these fields are never defined at the same integration time in a trajectory, which spoils the evaluation of the hidden constraint. In the case of SU(N) gauge fields, the situation is even worse. First of all, the definition of a gauge invariant Higgs field leads to composite fields in terms of the underlying gauge variables. Further, the equations of motion define the change of the fields in HMC algorithms not by an additive but a multiplicative exponential term proportional to the momenta $\pi(n_\mu)$, which causes an additional challenge for the determination of the Lagrange multiplier(s), cf. sections~\ref{sec:chmctor} and~\ref{sec:chmcorb} and appendices~\ref{app:chmctor}-\ref{app:chmcorb}. 

We use an extension of the Newton-St\"ormer-Verlet-leapfrog method, the so-called Rattle algorithm for general Hamiltonians of constrained systems~\cite{hairer:2002gni,hairer:2003gni}, with an additional half integration step for the momenta $\pi$ ($\pi_{n+1/2}$ to $\pi_{n+1}$, the index $n$ denotes the molecular dynamics time step $nh$, with the integration step size $h$). This ensures to have field and momentum variables at the same integration time and allows us to apply the hidden constraint. The implementations of the constrained equations of motion for our special cases using the Rattle algorithm are detailed in appendix~\ref{app:rattle}. In the next chapters we summarize the new algorithms for the various models with numerical tests of their time-reversibility. Further, we present first results for the constraint effective potentials and compare them to Higgs potentials from unconstrained simulations using the histogram method and a one-loop Higgs potential~\cite{Irges:2017ztc}. 

\section{4D ABELIAN-HIGGS MODEL}\label{sec:chmcabel}

The action of the 4D Abelian-Higgs model is given by
\bea
S[U_\mu,\phi]&=&S_g[U_\mu]+S_\phi[U_\mu,\phi], \quad S_g[U]=\beta\sum_{n_\mu}\sum_{\mu<\nu}\left\{1-\mbox{Re}U_{\mu\nu}(n_\mu)\right\}\\
S_\phi[U_\mu,\phi] &=& \sum_{n_\mu}
|\phi(n_\mu)|^2-2\kappa\sum_\mu\mbox{Re}\left\{\phi^\dagger(n_\mu)[U_\mu(n_\mu)]^q\phi(n_\mu+a\hat\mu)\right\}+\lambda(|\phi(n_\mu)|^2-1)^2\label{eq:Sabel}
\eea
with $\beta$ and $\lambda$ the gauge and quartic couplings, respectively, $\kappa$ the hopping (mass) parameter, $\phi=\phi_1+i\phi_2$ a complex scalar field, $U_\mu(n_\mu)$ U(1) gauge links and $U_{\mu\nu}(n_\mu)=U_\mu(n_\mu)U_\nu(n_\mu+\hat\mu)U_\mu^\dagger(n_\mu+\hat\nu)U_\nu^\dagger(n_\mu)$ the standard plaquettes.
$n_\mu$ ($\mu=0,1,2,3$) are the integer coordinates of the points on the 4D lattice of volume $\Omega=L_s^3\times L_t$ and we use a charge parameter $q=1$. 

\subsection{Constrained simulation}
In order to respect gauge invariance of the 4D Abelian-Higgs model, the (composite) Higgs field is constructed via $\mathcal H(n_\mu)=\phi^\dagger(n_\mu)\phi(n_\mu)$ and our constraint condition reads 
\bea\label{eq:constraint}
\dfrac{1}{\Omega}\sum_{n_\mu}\phi^\dagger(n_\mu)\phi(n_\mu)=\dfrac{1}{\Omega}\sum_{n_\mu,i=1,2}\phi_i(n_\mu)^2=\Phi.
\eea 
which has to be fulfilled at all times, therefore, the field variables $\phi(n_\mu)$ have to be initialized with respect to the constraint already. The hidden constraint is given by the first derivative of the constraint condition with respect to integration time, {\it i.e.},  
\bea\label{eq:hiddenconstraint}
\sum_{n_\mu,i=1,2}\phi_i(n_\mu)\dot\phi_i(n_\mu)=\sum_{n_\mu,i=1,2}\phi_i(n_\mu)\pi_i(n_\mu)=0,
\eea 
which has to vanish in order for the constraint condition to be fulfilled at all times. Therefore, when drawing the Gaussian-distributed random conjugate momenta $\pi^r(n_\mu)$ we have to ensure that they comply with the hidden constraint \eq{eq:hiddenconstraint}, which we achieve via orthogonal projection~\cite{hairer:2002gni}
\bea
\pi_{i,0}(n_\mu)=\pi_i^r(n_\mu)-\dfrac{\phi_i(n_\mu)}{\Omega\Phi}\sum_{m_\mu,j=1,2}\pi_j^r(m_\mu)\phi_j(m_\mu).
\eea 

\noindent $\pi_{i,0}$ are defined as a linear transformation of $\{\pi_j^r\}$ and, therefore, are still normally distributed around zero.
The constrained HMC algorithm for the Abelian-Higgs model can be formulated in the following way, using the so-called Rattle algorithm~\cite{hairer:2002gni,hairer:2003gni} (see appendix~\ref{app:abel} for the derivation)
\bsa
\pi_{i,n+1/2}&=&\pi_{i,n}-\dfrac{h}{2}\bigg(\dfrac{\partial S}{\partial\phi_{i,n}}+\dfrac{2\phi_{i,n}\lambda_n^{(1)}}{\Omega}\bigg)\;,
\qquad P_{\mu,n+1/2}\;=\;P_{\mu, n}-\dfrac{h}{2}\dfrac{\partial S}{\partial U_{\mu, n}}\label{eq:rattleAHa}\\
\phi_{i,n+1}&=&\phi_{i,n}+h\pi_{i,n+1/2}\;,\qquad\qquad\qquad\qquad U_{\mu, n+1}\;=\;U_n+hP_{\mu, n+1/2}\label{eq:rattleAHb}\\
\lambda_n^{(1)}&=&\dfrac{\Omega}{h^2}-\sum_{n_\mu,i}\dfrac{\phi_{i,n}}{2\Phi}\dfrac{\partial S}{\partial\phi_{i,n}}\pm\sqrt{\dfrac{\Omega^2}{h^4}+\bigg(\sum_{n_\mu,i}\dfrac{\phi_{i,n}}{2\Phi}\dfrac{\partial S}{\partial\phi_{i,n}}\bigg)^2-\dfrac{\Omega}{\Phi}\sum_{n_\mu,i}\bigg(\dfrac{\pi_{i,n}}{h}-\dfrac{1}{2}\dfrac{\partial S}{\partial\phi_{i,n}}\bigg)^2}\label{eq:lambel}\\
\pi_{i,n+1}&=&\pi_{i,n+1/2}-\dfrac{h}{2}\bigg(\dfrac{\partial S}{\partial\phi_{i,n+1}}+\dfrac{2\phi_{i,n+1}\lambda_n^{(2)}}{\Omega}\bigg)\label{eq:rattleAHd}\\
\lambda_n^{(2)}&=&\sum_{n_\mu,i}\bigg(\dfrac{\phi_{i,n+1}\pi_{i,n+1/2}}{h\Phi}-\dfrac{\phi_{i,n+1}}{2\Phi}\dfrac{\partial S}{\partial\phi_{i,n+1}}\bigg)
\label{eq:rattleAHe}
\esa
where $X_{i,n}\equiv X_{i,n}(n_\mu)$ at molecular dynamics (MD) time $nh$ with the (MD) integration step size $h$. 
The gauge links $U_\mu(n_\mu)$ and corresponding conjugate momenta $P_\mu(n_\mu)$ are updated using the standard leap-frog algorithm. For the Higgs field $\phi(n_\mu)$ and conjugate momenta $\pi(n_\mu)$ the first three (left) equations (\ref{eq:rattleAHa}-\ref{eq:lambel}) determine $\pi_{n+1/2}$ and $\phi_{n+1}$, such that the constraint is fulfilled at integration step $n+1$. During numerical simulations it turns out that only the $-$ sign in front of the square root fulfills the constraint condition. Equations (\ref{eq:rattleAHd}-\ref{eq:rattleAHe}) ensure the hidden constraint for fields $\phi_{n+1}$ and momenta $\pi_{n+1}$ at the same integration time, before starting over, {\it i.e.}, continuing to integration times $n+3/2$ and $n+2$ subsequently.

\medskip

We check numerically the time reversibility by performing one trajectory with stepsize $+h$ and another one with $-h$, retrieving the initial field and momentum variables. Further, we calculate the Jacobian $J=\dfrac{\partial(\phi_{n+1}(n_\mu),\pi_{n+1}(n_\mu))}{\partial(\phi_n(m_\mu),\pi_n(m_\mu))}$ numerically, yielding a $(4^{L^4})^2$ matrix with det$J=1$, implying volume preservation. This is just a test of our implementation since the Rattle algorithm ensures these two and other necessary geometric properties, see appendix~\ref{app:rattle}.

\subsection{Constraint effective potential}

Using the algorithm we want to measure the derivative of the constraint effective potential $U_{\Omega,\rm{cnst.}}'(\Phi)=-\dfrac{1}{\Omega}\big\langle\lambda^{(1)}\big\rangle_\Phi$ during Monte Carlo simulations. The numerical observable $\lambda^{(1)}$ however, depends on the molecular dynamics integration stepsize $h$, which is not a physical quantity and, therefore, we want to analyze the continuum limit $h\rightarrow0$ of this observable by rewriting the square root as a Taylor series 
\bea
\lambda^{(1)}
&\stackrel{h\rightarrow0}{=}&\dfrac{1}{2\Phi}\sum_{n_\mu,i}\bigg(\pi_i^2-\phi_i\dfrac{\partial S}{\partial\phi_i}\bigg)\;\Rightarrow\;U_{\Omega,\rm{cnst.}}'\equiv\dfrac{1}{2\Omega\Phi}\bigg\langle\sum_{n_\mu,i}\bigg(\phi_i\dfrac{\partial S}{\partial\phi_i}-\pi_i^2\bigg)\bigg\rangle_\Phi\label{eq:contAH}
\eea

\begin{figure}[h]
\includegraphics[width=.5\linewidth]{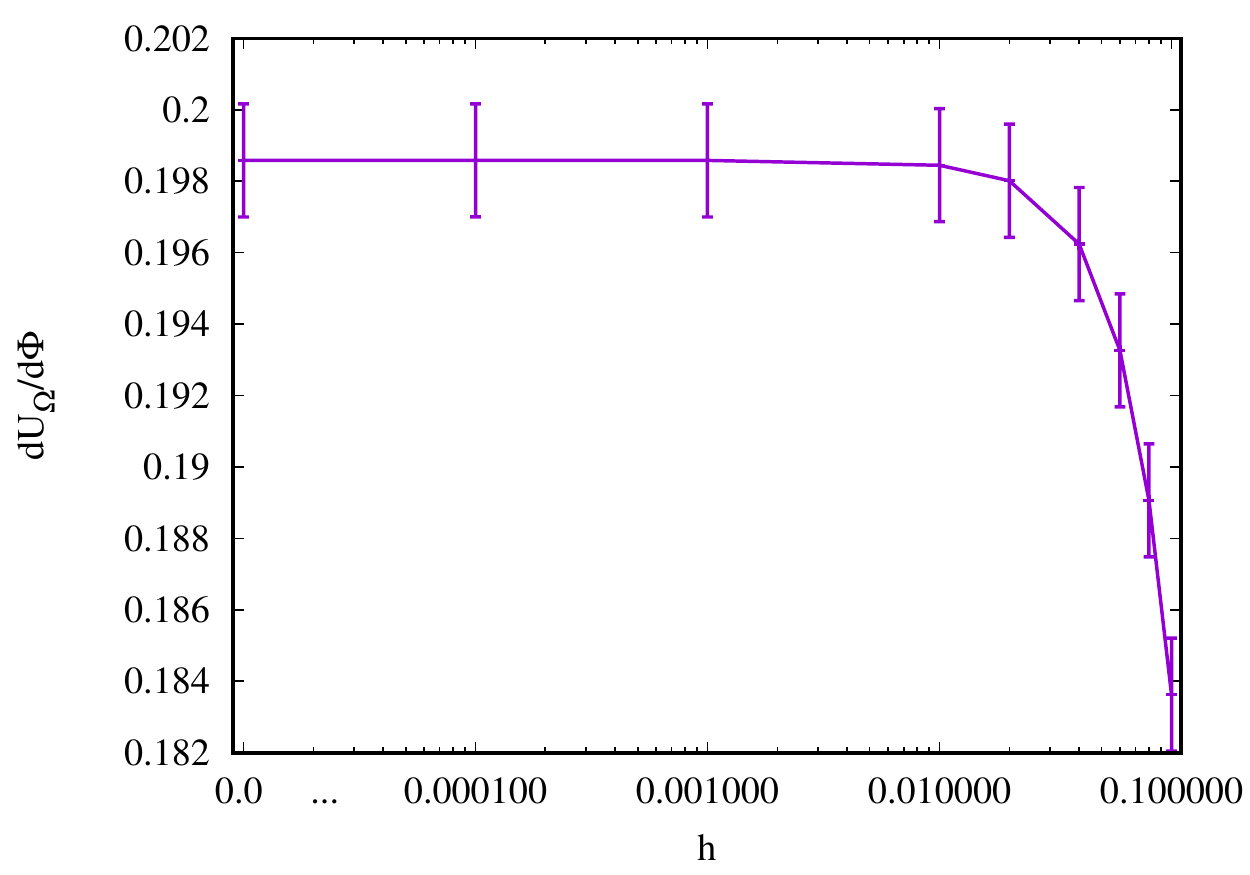}
\includegraphics[width=.5\linewidth]{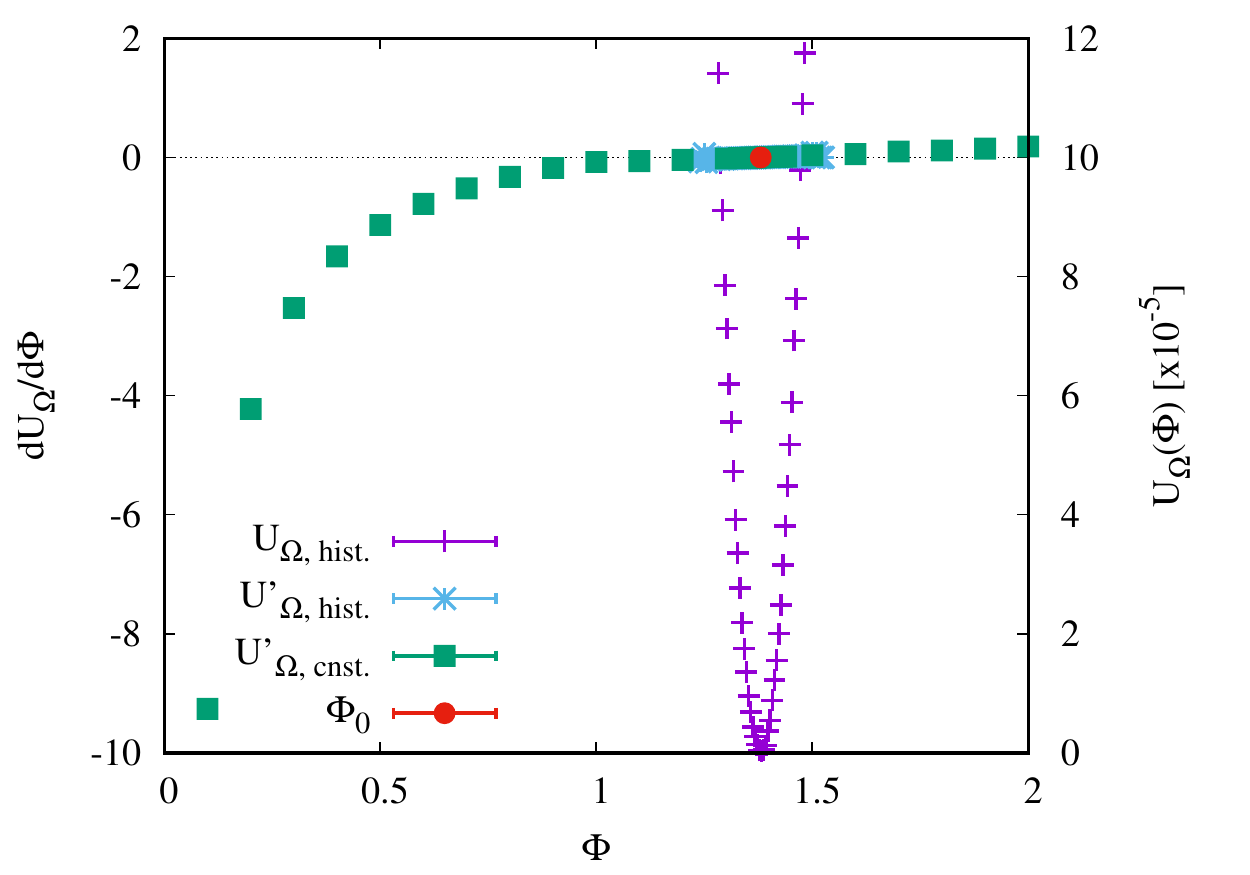}
\caption{(left) Well defined continuum limit of $U_{\Omega,\rm{cnst.}}'$ (\ref{eq:numAH}) with respect to step size $h\rightarrow0$, it agrees with the continuum form $U_{\Omega,\rm{cont.}}'$ (\ref{eq:contAH}) at $h=0$ for $h\leq0.01$. (right) Effective potential and derivatives from histogram method and constrained simulations for the Abelian-Higgs model at $\beta=1.4, \kappa=0.17, \lambda=0.15$ on $\Omega=8^4$ lattices. The potential and its derivative diverge for $\Phi\rightarrow0$.}\label{fig:contAH}
\end{figure}

In \fig{fig:contAH} (left) we investigate the continuum limit $h\rightarrow0$ by plotting $U_{\Omega,\rm{cnst.}}'$ for various simulation step sizes $h$, rapidly approaching the continuum value $U_{\Omega,\rm{cont.}}'$ at $h=0$. We conclude that for the purpose of measuring the effective potential an integration step size of $h\leq0.01$ is sufficient, for simplicity however, we use the continuum form anyhow. 

\begin{figure}[h]
\includegraphics[width=.5\linewidth]{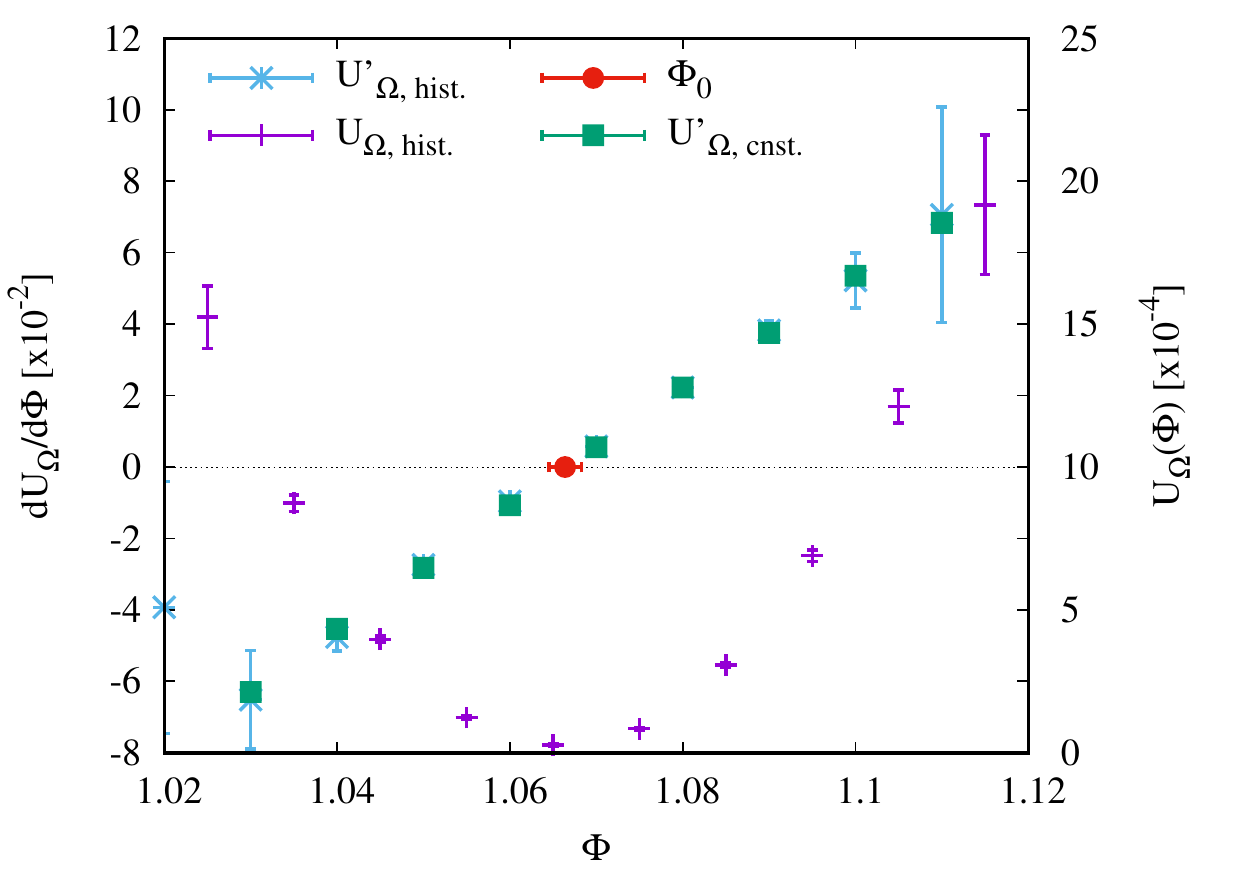}
\includegraphics[width=.5\linewidth]{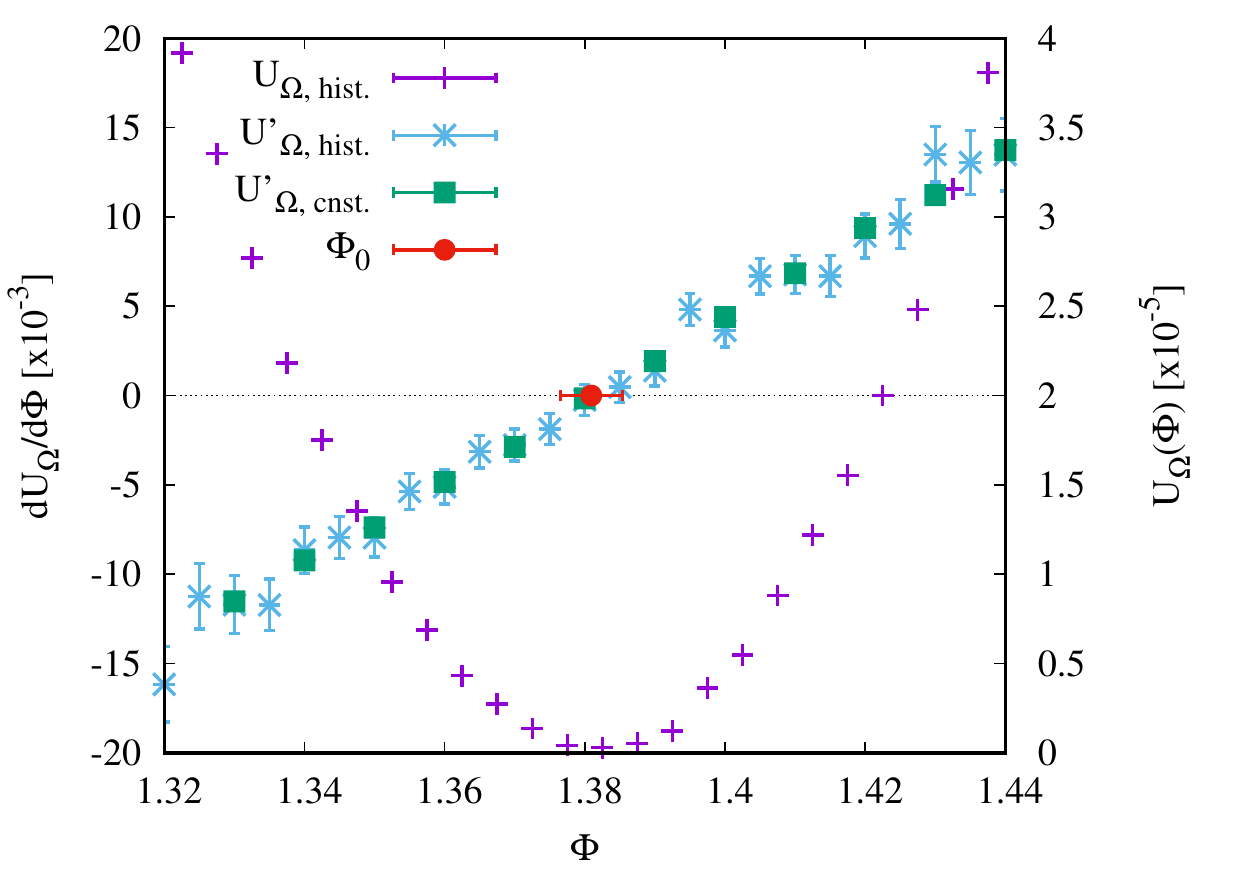}
\caption{Effective potential and derivatives from histogram method and constrained simulations for the Abelian-Higgs model at $\beta=0.6, \kappa=0.3, \lambda=1$ (left) and $\beta=1.4, \kappa=0.17, \lambda=0.15$ (right, zoom of right plot in \fig{fig:contAH}) on $\Omega=8^4$ lattices.}\label{fig:UeffAH}
\end{figure}

First results of the effective potential in the Higgs phase are presented in \fig{fig:contAH} (right) and \fig{fig:UeffAH} for $\beta=1.4, \kappa=0.17, \lambda=0.15$ and $\beta=0.6, \kappa=0.3, \lambda=1$ on $\Omega=8^4$ lattices, comparing the derivative of the constraint effective potential $U_{\Omega,\rm{cnst.}}'$ with the effective potential $U_{\Omega,\rm{hist.}}$ and its derivative obtained from a standard histogram method, {\it i.e.}, measuring the distribution of the field $\Phi=\sum_{n_\mu}\phi(n_\mu)^\dagger\phi(n_\mu)$ in an unconstrained simulation, appropriately binning it in a normalized histogram and taking the logarithm. The unconstrained simulation for the histogram method needs much more statistics than the individual constrained simulations combined to achieve comparable precision, only in the vicinity of the expectation value of the Higgs field $\Phi_0=\langle\Phi\rangle$. Note that the latter exactly coincides with the zero crossing of the derivative of the (constraint) effective potential, and we can read off the Higgs mass from the second derivative of the (constraint) effective potential at $\Phi_0$. Further notice in the right plot of \fig{fig:contAH} that with the new method to measure the derivative of the constraint effective potential, we can access the Higgs potential over the full parameter range of $\Phi$ with very high precision and find in the case of the Abelian-Higgs model that it diverges for $\Phi\rightarrow0$, since only positive values of $\Phi$ are allowed by definition, see Eq. (\ref{eq:constraint}).  

\subsection{Comparison to the one-loop Higgs potential in unitary gauge}\label{sec:unit}

Note, we can write the action~(\ref{eq:Sabel}) in unitary gauge using the variable transformation proposed in \cite{Montvay:1994cy} p.322, $\phi(n_\mu)=\rho(n_\mu)\exp{i\varphi(n_\mu)}\;\Rightarrow\;\phi_1=\rho\cos\varphi,\;\phi_2=\rho\sin\varphi$:
\vspace{-4mm}
\bean
S_\rho[V_\mu,\rho]&=&\sum_{n_\mu}\bigg[\rho(n_\mu)^2+\lambda(\rho(n_\mu)^2-1)^2-2\kappa\rho(n_\mu)\mbox{Re}\sum_\mu\rho(n_\mu+\hat\mu)\overbrace{e^{-i\varphi(n_\mu+\hat\mu)}U_{\mu}(n_\mu)e^{i\varphi(n_\mu)}}^{=V_{\mu}(n_\mu)}\bigg],
\ean
with gauge invariant links $V_\mu(n_\mu)$. This allows us to rewrite the constrained Hamiltonian as\footnote{Notice, that we always use the same symbol $\Phi$ to denote different constraint fields.}
\bean
&&\tilde H[V_\mu,\rho]=S_\rho[V_\mu,\rho]-\sum_{n_\mu}\ln[\rho(n_\mu)]+\dfrac{1}{2}\sum_{n_\mu}\pi(n_\mu)^2+\lambda^{(1)}\bigg(\dfrac{1}{\Omega}\sum_{n_\mu}\rho(n_\mu)-\Phi\bigg),
\ean
where the $\ln(\rho)$ term enters from the Jacobian of the variable transformation and plays an important role: for small $\rho$ we get a diverging contribution to the action which pushes the system away from $\rho\leq0$ which would be unphysical. 
The constrained equations of motion read
\bean
\dot\rho(n_\mu)=\;\;\dfrac{\partial\tilde H}{\partial\pi(n_\mu)}\,=\;\pi(n_\mu)
\;,\qquad
\dot\pi(n_\mu)=-\dfrac{\partial\tilde H}{\partial\rho(n_\mu)}=-\dfrac{\partial S_\rho}{\partial\rho(n_\mu)}+\dfrac{1}{\rho(n_\mu)}-\dfrac{\lambda^{(1)}}{\Omega}
\ean
and derivatives of the constraint condition with respect to the molecular dynamics time allow us to solve for the Lagrange multiplier $\lambda^{(1)}$ 
\bean
&&\sum_{n_\mu}\dot\rho(n_\mu)=\sum_{n_\mu}\pi(n_\mu)=0\Rightarrow\sum_{n_\mu}\dot\pi(n_\mu)=0\Rightarrow\lambda^{(1)}=\sum_{n_\mu}\bigg(\dfrac{1}{\rho(n_\mu)}-\dfrac{\partial S_\rho}{\partial\rho(n_\mu)}\bigg).
\ean
We can use the standard leap-frog algorithm to perform the HMC updates as shown in~\cite{Fodor:2007fn} for a Higgs-Yukawa theory with $N_f$ fermions. In order to guarantee that the hidden constraint is fulfilled by the algorithm (note that the leap-frog algorithm would yield momenta fulfilling the constraint at the new time point, as it preserves linear constraints only in the momenta exactly), one has to initialize the (random) fictitious momenta $\pi(n_\mu)$ in each trajectory accordingly, {\it i.e.}, with respect to $\sum_{n_\mu}\pi(n_\mu)=0$. During the constrained simulations we measure the derivative of the effective potential\footnote{In the proceedings~\cite{Hollwieser:2018nrh} the derivative of the constraint effective potential in the 4D Abelian-Higgs model in unitary gauge was missing a contribution and, therefore, the result presented in Fig. 1 of the proceedings is inaccurate.} 
$U'_\Omega(\Phi)=-\frac{1}{\Omega}\langle\lambda^{(1)}\rangle=\frac{1}{\Omega}\langle\sum_{n_\mu}[\partial S_\rho/\partial\rho(n_\mu)-1/\rho(n_\mu)]\rangle_\Phi$, 
where $\langle\ldots\rangle_\Phi$ means the expectation value at fixed $\Phi=\Omega^{-1}\sum_{n_\mu}\rho(n_\mu)$.  

\begin{figure}[h!]
\centering
$\qquad\quad\beta=8, \kappa=0.166, \lambda=0.15, \Omega=4^4\qquad\qquad\qquad\;\beta=8, \kappa=0.166, \lambda=0.15, \Omega=8^4\qquad\quad$
\includegraphics[width=0.495\linewidth]{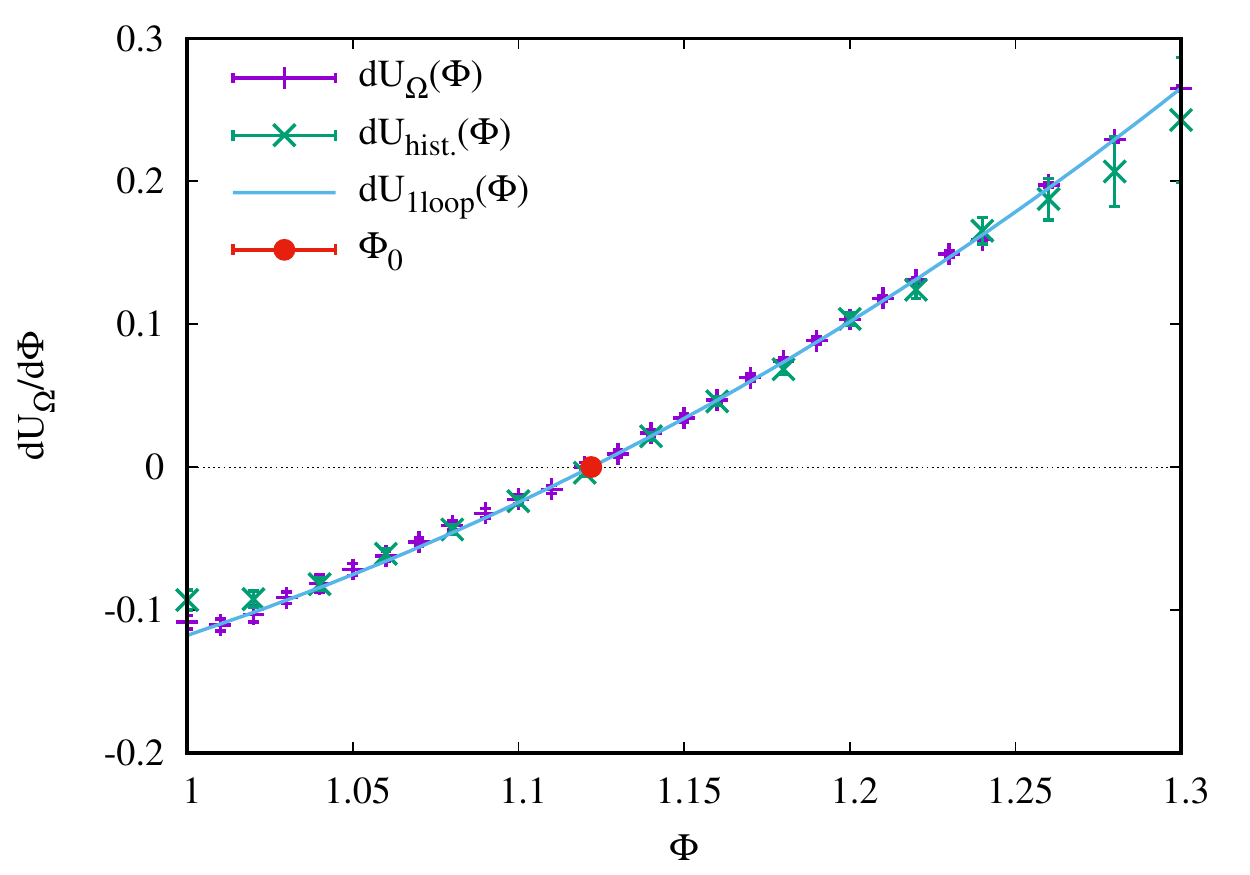}
\includegraphics[width=0.495\linewidth]{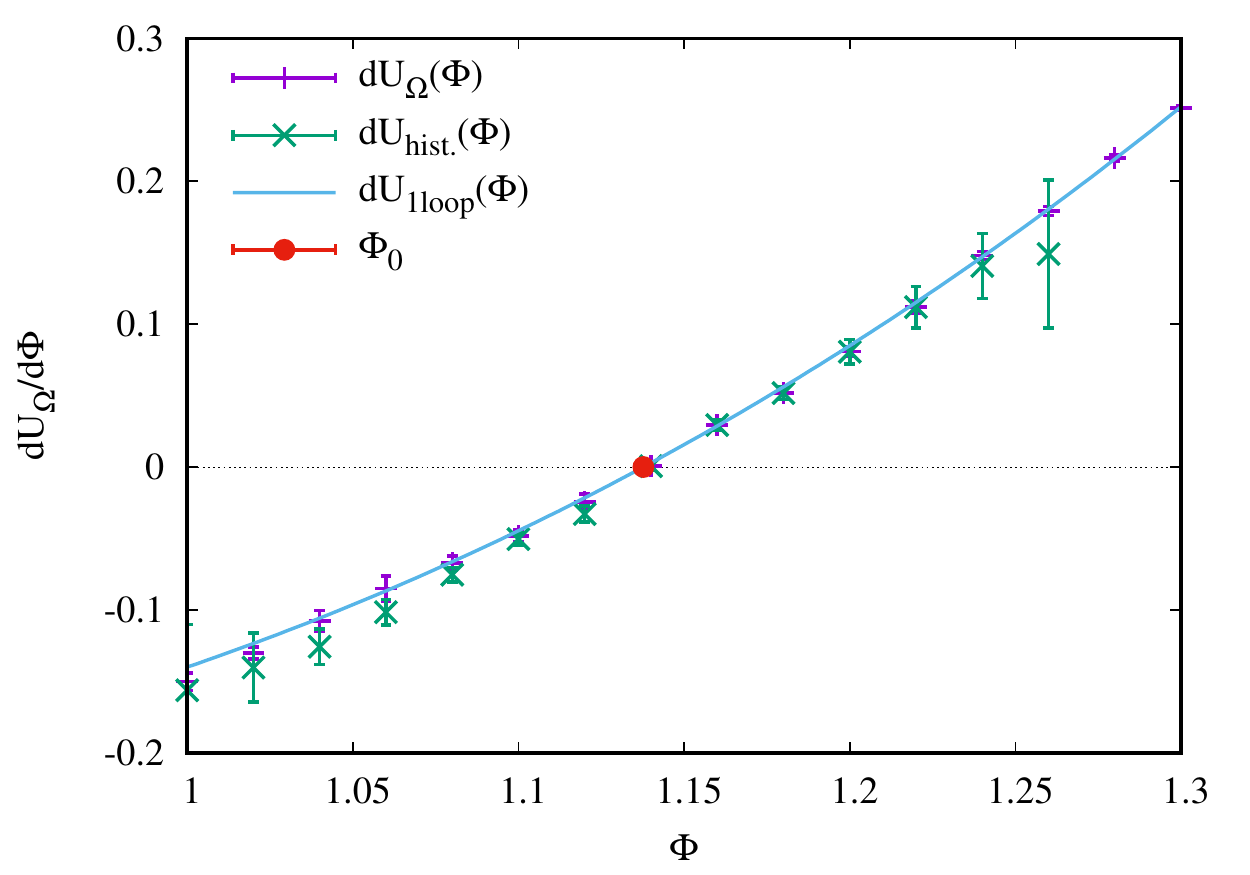}\\
$\beta=8, \kappa=0.166, \lambda=0.15, \Omega=8^3\times16$\\
\includegraphics[width=0.495\linewidth]{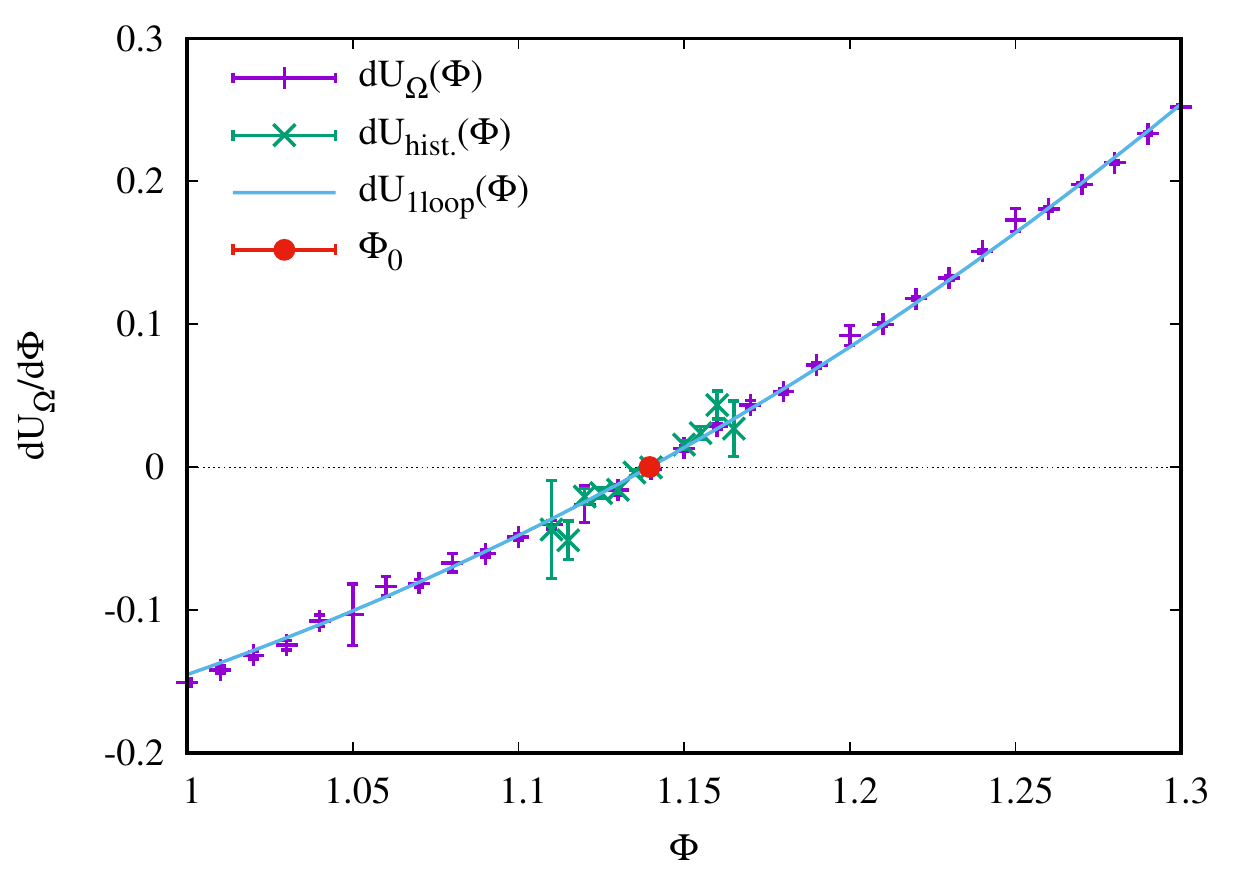}
\includegraphics[width=0.495\linewidth]{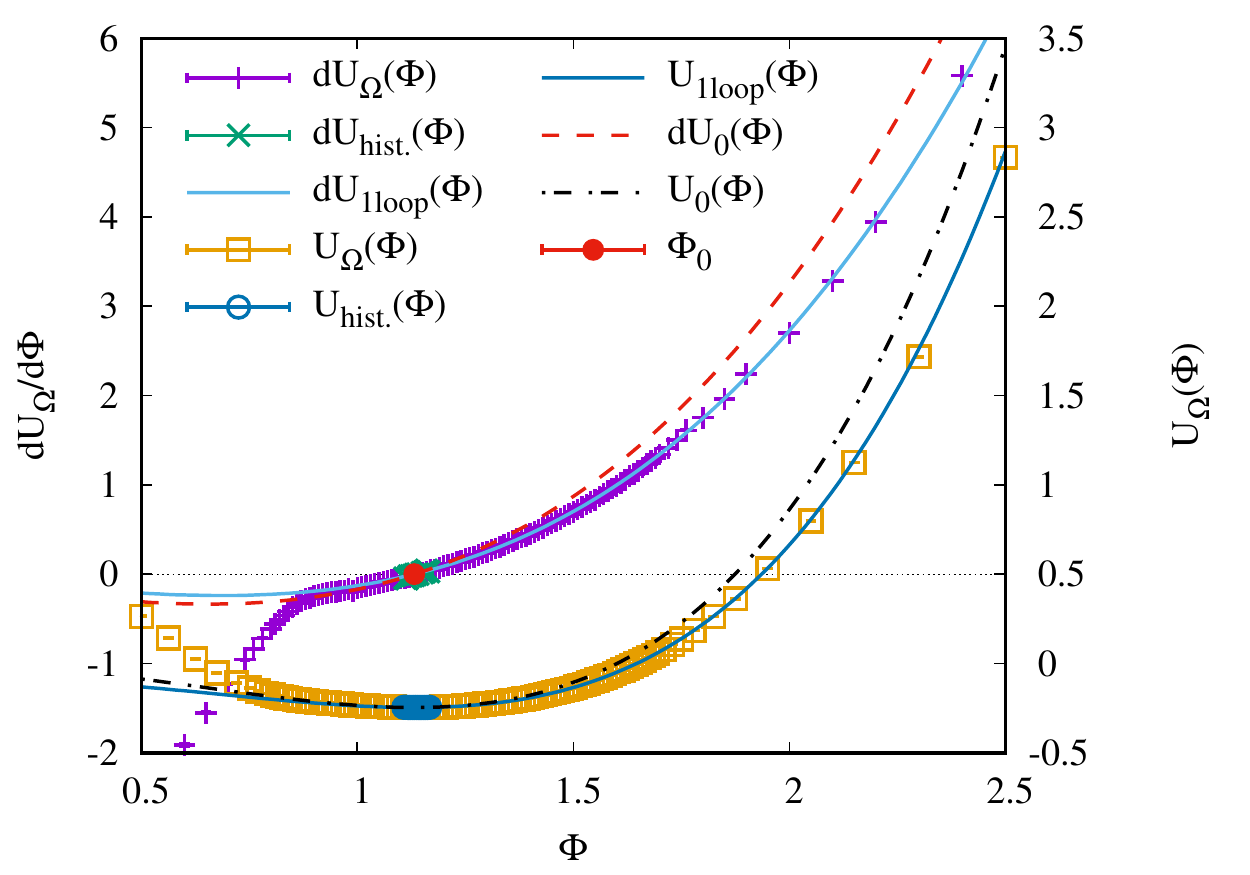}
\vspace{-3mm}
\caption{Derivatives of the effective potential from histogram method and constrained simulations for the Abelian-Higgs model in unitary gauge at $\beta=8, \kappa=0.166$ and $\lambda=0.15$ and different lattice volumes $\Omega$, fitted with the derivative of the one-loop Higgs potential $V_1(\phi)$~\cite{Irges:2017ztc} via $m_H$. The fits work quite well in the vicinity of $\Phi_0$, allowing us to extract reasonable Higgs masses, see also \tab{tab:mass} and~\ref{tab:masV}. The plot on the bottom to the right is an overview plot to the one on the left, showing the effects of the one-loop corrections compared to the classical potential $U_0(\Phi)=-m_H^2\Phi^2/2+\lambda\Phi^4$, where $m_H=2\Phi_0\sqrt{\lambda}$, also listed in~\tab{tab:mass}. In order to plot the correct classical potential, the histogram, constraint and one-loop potentials are shifted by an integration constant $U_0(\Phi_0)$ in the bottom right plot.}\label{fig:uniteffS1}
\end{figure}

Results are presented in \fig{fig:uniteffS1}, where we compare the constraint effective potential with the effective potential measured by the histogram method in unconstrained simulations and with the 
finite, one-loop Higgs potential given by~\cite{Irges:2017ztc} (here $\tilde\lambda=4\lambda$)
\bea\label{eq:1loopV}
V_1(\phi) = \frac{1}{2} m_H^2 \phi^2 + \left[ \sqrt{\frac{\tilde\lambda}{2}} m_H - \frac{m_H}{16 \pi^2\sqrt{2\tilde\lambda}} \Bigg ( 
9 \tilde\lambda^2 + \frac{8 \tilde\lambda^2 m_Z^4}{ m_H^4} \Bigg ) \right]{\phi ^3} 
+\frac{\phi^4}{4}\left[ \tilde\lambda - \frac{1}{16 \pi^2} \Bigg ( \frac{32 \tilde\lambda^2 m_Z^4}{ m_H^4} \Bigg ) \right]
\eea
via fitting the (bare) Higgs mass $m_H$. We actually fit the derivative $U'_{1\rm{loop}}(\Phi)=V'_1(\Phi-\Phi_0)$ to our measured $U'_\Omega(\Phi)$, using the bare quartic coupling $\lambda$ and Z-boson mass given by the quasi-classical perturbative relation $m_Z=\sqrt{2\kappa g^2\langle\rho^2\rangle}$~\cite{Evertz:1986ur}, with the gauge coupling $g^2=1/\beta$. We choose a large value $\beta=8$ in order to stay in the weak coupling regime where we expect renormalization effects to be small. We find that the one-loop formula fits the constraint potential much better than the classical ansatz $U_0(\Phi)=-m_H^2\Phi^2/2+\lambda\Phi^4$, while the histogram data cannot differentiate the one-loop corrections within their limited range of $\Phi$.

During the unconstrained simulations we also measure the two-point function~\cite{Evertz:1986ur}
\bea
C_i(\Delta t)=L_t^{-1}\sum_t\langle(O_i(t)-\langle O_i(t)\rangle)(O_i(t+\Delta t)-\langle O_i(t+\Delta t)\rangle)\rangle\approx\mathrm{cnst}(e^{-m\Delta t}+e^{-m(L_t-\Delta t)})\label{eq:corr}
\eea
of the following lattice operators associated with quantum numbers $J^{PC}= 0^{++}$ and $1^{--}$
\bea
O_H(t)&=&\Omega_3^{-1}\,\mathrm{Re}\sum_x\sum_{\mu=1,2,3}\rho(x,t)V_\mu(x,t)\rho(x+\hat\mu,t)\\
O_Z(t)&=&\Omega_3^{-1}\,\mathrm{Im}\sum_x\sum_{\mu=1,2,3}\rho(x,t)V_\mu(x,t)\rho(x+\hat\mu,t)
\label{eq:ops}
\eea 
with the spatial (3D) volume $\Omega_3$. Fitting with the ansatz given on the right hand side of \eq{eq:corr}, discarding points with $\Delta t=0$ and $1$, we can extract the renormalized Higgs mass $m_{H,R}$ from the first operator $O_H$ and the gauge Z-boson (massive photon) mass $m_{Z,R}$ from $O_Z$
, the determined masses are summarized in \tab{tab:mass}. It can be seen, that renormalization effects are indeed small for $\beta=8$ (in contrast to $\beta=2.5$) and the masses agree quite well. We also list the classical values of the Higgs mass $2\Phi_0\sqrt{\lambda}$, derived from the classical ansatz $U_0(\Phi)$, which quantify the discrepancy of the classical Higgs potential ansatz, wich is just the tree level approximation, compared to the one-loop result.

\begin{table}[h!]
\begin{center}
\begin{tabular}{l|l|l|l|l|lll|ll}
$\beta$ & $\lambda$ & $\kappa$ & $\Phi_0$ & $\langle\rho^2\rangle$ & $2\Phi_0\sqrt{\lambda}$ & $m_H$ & $m_{H,R}$ & $m_Z$ & $m_{Z,R}$ \\
\midrule
2.5 & 3.0 & 0.184 & 0.931(1) & 0.913(2) & 3.229(1) & 4.071(1) & 0.662(19) & 0.362(1) & 0.276(17) \\ 
\hline
8 & 0.15 & 0.164 & 1.100(1)& 1.414(2) & 0.852(1) & 1.062(1) & 1.054(19) & 0.241(1) & 0.176(13) \\
& & 0.166 & 1.133(1) & 1.489(1) & 0.878(1) & 1.093(1) & 1.099(21) & 0.247(1) & 0.185(12) \\
& & 0.168  & 1.164(1) & 1.563(1) & 0.902(1) & 1.131(1) & 1.162(23) &  0.256(1) & 0.209(14) \\
& & 0.17 & 1.194(1) & 1.634(3) & 0.925(1) & 1.162(1) & 1.218(28) & 0.263(1) & 0.224(17) \\
& & 0.2 & 1.555(1) & 2.593(2) & 1.204(1) & 1.575(1) & 1.787(26) & 0.360(1) & 0.328(19) \\
\end{tabular}
\end{center}
\vspace{-3mm}
\caption{Higgs and Z-boson masses from fits of the one-loop potential \eq{eq:1loopV} to the constraint effective potential on  $\Omega=8^3\times16$ volumes via $m_H$ using the quasi-classical relation $m_Z=\sqrt{2\kappa g^2\langle\rho^2\rangle}$~\cite{Evertz:1986ur}, and from fits of the two-point function \eq{eq:corr} using the operators given in \eq{eq:ops} for $m_{H,R}$ and $m_{Z,R}$. We also list the classical values of the Higgs mass $2\Phi_0\sqrt{\lambda}$ quantifying the discrepancy of the classical potential compared to the one-loop result.}
\label{tab:mass}
\end{table}

In \tab{tab:masV} we show that the extracted Higgs mass does not depend on the volume and the precision of the new method does not deteriorate when increasing the latter, in contrast to the histogram method.
We conclude that the constraint effective potential accurately determines the Higgs mass and reproduces not only the effective potential from the histogram method, but also compares very well to the one-loop Higgs potential given in \eq{eq:1loopV}~\cite{Irges:2017ztc}.

\begin{table}[h!]
\begin{center}
\begin{tabular}{l|l|l|l|ll|ll|l}
volume & $\Phi_0$ & $m_H$ & $U'_\Omega(1.0)$ & $U'_\Omega(1.11)$ & $U'_\mathrm{h.}(1.11)$ & $U'_\Omega(1.15)$ & $U'_\mathrm{h.}(1.15)$  & $U'_\Omega(1.3)$\\ 
\midrule
$4^4$ & 1.122(8) & 1.068(1) & -0.099(3) & -0.016(3) & -0.014(3) & 0.033(3) & 0.034(3) & 0.264(2) \\
$4^3\times16$ & 1.132(1) & 1.092(1) & -0.142(2) & -0.032(3) & -0.019(8) & 0.018(3) & 0.025(5) & 0.261(3) \\
$8^4$ & 1.133(1) & 1.093(1) & -0.159(2) & -0.040(2) & -0.043(17) & 0.015(3) & 0.017(9) & 0.250(1) \\
$8^3\times16$ & 1.133(1) & 1.093(1) & -0.160(2) & -0.040(3) & -0.044(34) & 0.013(3) & 0.016(13) & 0.252(1)
\end{tabular}
\end{center}
\vspace{-3mm}
\caption{Volume scaling of the Higgs mass $m_H$ from fits to the effective potential, except for the smallest volume we don't see an effect, and comparison of the precision of the derivative of the potential obtained from constrained ($U_\Omega'$) and unconstrained (histogram, $U_{\rm h.}'$) simulations. Contrary to the latter, the error of results from constrained simulations does not increase with the volume. The results are for $\beta=8, \kappa=0.166$ and $\lambda=0.15$.}
\label{tab:masV}
\end{table}

\section{5D SU(2) GAUGE THEORY ON THE TORUS}\label{sec:chmctor}

The anisotropic Wilson plaquette action for a 5D SU(2) gauge theory with periodic (torus) boundary conditions is given by~\cite{Ejiri:2000fc,Knechtli:2011gq,Knechtli:2016pph}
\begin{equation}\label{eq:toraction}
S_W^{tor} = \sum_{n_\mu}\sum_{n_5=0}^{N_5-1}\bigg[\frac{\beta_4}{2}\sum_{\mu<\nu} \mathrm{Re}\,\tr\{1-U_{\mu\nu}(n_\mu,n_5)\} +
\frac{\beta_5}{2}\sum_{\mu}\mathrm{Re}\,\tr\{1-U_{\mu5}(n_\mu,n_5)\}\bigg] \,,
\end{equation}
where $\beta_4$ and $\beta_5$ are the gauge couplings associated with plaquettes
spanning the standard four dimensions ($U_{\mu\nu}$) and the fifth dimension ($U_{\mu5}$)
respectively. The anisotropy is $\gamma=\sqrt{\beta_5/\beta_4}$ and in the classical limit
$\gamma=a_4/a_5$, where $a_4$ denotes the lattice spacing in the usual
four dimensions and $a_5$ denotes the lattice spacing in the extra dimension.
The theory is defined on the periodic interval $I=\{n_\mu,0\le n_5 < N_5\}$, where
$(n_\mu,n_5)$, $\mu=0,1,2,3$ are the integer coordinates of the points of the five-dimensional lattice. 

We start with a Higgs field $\mathcal H(n_\mu)=\tr P_5(n_\mu)$ given by the Polyakov loops in the extra dimension $P_5(n_\mu)=\prod_{n_5=0}^{N_5-1}[U_5(n_\mu,n_5)]$, and the constraint condition is given by
$\frac{1}{2\Omega}\sum_{n_\mu}\tr P_5(n_\mu) =\Phi$ (\ref{eq:trPcnst}). Hence, only the links in the extra dimension will be affected by the constraint, all other links $U_\mu(n_\mu,n_5)$, $\mu=0,1,2,3$ can be updated using the standard leapfrog method. For the   links $U_5(n_\mu,n_5)$ and momenta $\pi_5(n_\mu,n_5)$ we apply the Rattle algorithm in appendix~\ref{app:chmctor} and find
\bsa
\pi_{n+1/2}&=&\pi_n-\dfrac{h}{2}\biggr(\dfrac{\partial S}{\partial U_n}-\dfrac{\lambda_n^{(1)}}{8\Omega}\tr[...\sigma_i U_n...]\sigma^i\biggr)\\
U_{n+1}&=&e^{h\pi_{n+1/2}}U_n\label{eq:trPupdt}\\
0&=&\dfrac{1}{2\Omega}\sum_{n_\mu}\tr P_{n+1}(n_\mu)-\Phi=\dfrac{1}{2\Omega}\sum_{n_\mu}\tr\prod_{n_5=0}^{N_5-1} U_{n+1}-\Phi\label{eq:trPcnst}\\
\pi_{n+1}&=&\pi_{n+1/2}-\dfrac{h}{2}\biggr(\dfrac{\partial S}{\partial U_{n+1}}-\dfrac{\lambda_n^{(2)}}{8\Omega}\tr[...\sigma_i U_{n+1}...]\sigma^i\biggr)\\
0&=&\dfrac{1}{8\Omega}\sum_{n_\mu,n_5}\tr\{\tr[...\sigma_i U_{n+1}...]\sigma^i\pi_{n+1}\} \label{eq:torcnst2a}
\esa

\noindent The term $\tr[...\sigma_iU_n...]\sigma^i$ denotes a Polyakov line at $n_\mu$ with an insertion of $\sigma_i$ at $n_5$, summing over $i=1,2,3$ for the three Pauli matrices. The first three equations determine $(\pi_{n+1/2},U_{n+1},\lambda_n^{(1)})$, whereas the remaining two give $(\pi_{n+1},\lambda_n^{(2)})$.

We  use a simple Secant method to get $\lambda_n^{(1)}$ up to machine precision, providing a precise root for the functional given by our constraint condition in \eq{eq:trPcnst}
\bean
f(\lambda_n^{(1)})&=&\dfrac{1}{2\Omega}\sum_{n_\mu}\tr P_{n+1}(n_\mu,\lambda_n^{(1)})-\Phi=\dfrac{1}{2\Omega}\sum_{n_\mu}\tr\prod_{n_5=0}^{N_5-1} U_{n+1}(n_\mu,n_5,\lambda_n^{(1)})-\Phi
\ean
with $U_{n+1}(n_\mu,n_5,\lambda_n^{(1)})$ given in \eq{eq:trPupdt}. We iterate $\lambda^{(1)}_{n,k+1}=\lambda^{(1)}_{n,k}-f(\lambda^{(1)}_{n,k})[\lambda^{(1)}_{n,k}-\lambda^{(1)}_{n,k-1}]/[f(\lambda^{(1)}_{n,k})-f(\lambda^{(1)}_{n,k-1})]$,
starting from an approximate solution $\lambda^{(1)}_{n,0}$
obtained by truncating the
the exponential in (\ref{eq:trPupdt}) after $\mathcal{O}(h^2)$
\bea
\dfrac{\lambda^{(1)}_{n,0}}{8\Omega}&=&\bigg\{\sum_{n_\mu,n_5}\bigg(\tr[...\dfrac{\partial S}{\partial U_n(n_\mu,n_5)}U_n(n_\mu,n_5)...]-\tr[...\pi_n^2(n_\mu,n_5)U_n(n_\mu,n_5)...]\nn
&&-2\sum_{m_5>n_5}^{N_5-1}\tr[...\pi_n(n_\mu,n_5)U_n(n_\mu,n_5)...\pi_n(n_\mu,m_5)U_n(n_\mu,m_5)...]\bigg)\bigg\}/\label{eq:lambda}\\
&&\qquad\sum_{n_\mu,n_5}\tr\{...\tr[...\sigma_i U_n(n_\mu,n_5)...]\sigma^iU_n(n_\mu,n_5)...\}\non
\eea
The iteration stops when $\lambda^{(1)}_{n,k+1}=\lambda^{(1)}_{n,k}$ or $f(\lambda^{(1)}_{n,k})=f(\lambda^{(1)}_{n,k-1})$ up to machine precision.

The second Lagrange multiplier is determined as (see appendix~\ref{app:chmctor} for details)
\bea
\dfrac{\lambda_n^{(2)}}{8\Omega}&=&\dfrac{\sum_{n_\mu,n_5}\tr[...\sigma_iU_{n+1}(n_\mu,n_5)...]\tr[\sigma^i\partial S/\partial U_{n+1}(n_\mu,n_5)-2\sigma^i\pi_{n+1/2}(n_\mu,n_5)/h]}{\sum_{n_\mu,n_5}\tr\{(\tr[...\sigma_i U_{n+1}(n_\mu,n_5)...]\sigma^i)^2\}}
\label{eq:mu}
\eea

\begin{figure}[h]
\begin{overpic}[width=.5\textwidth]{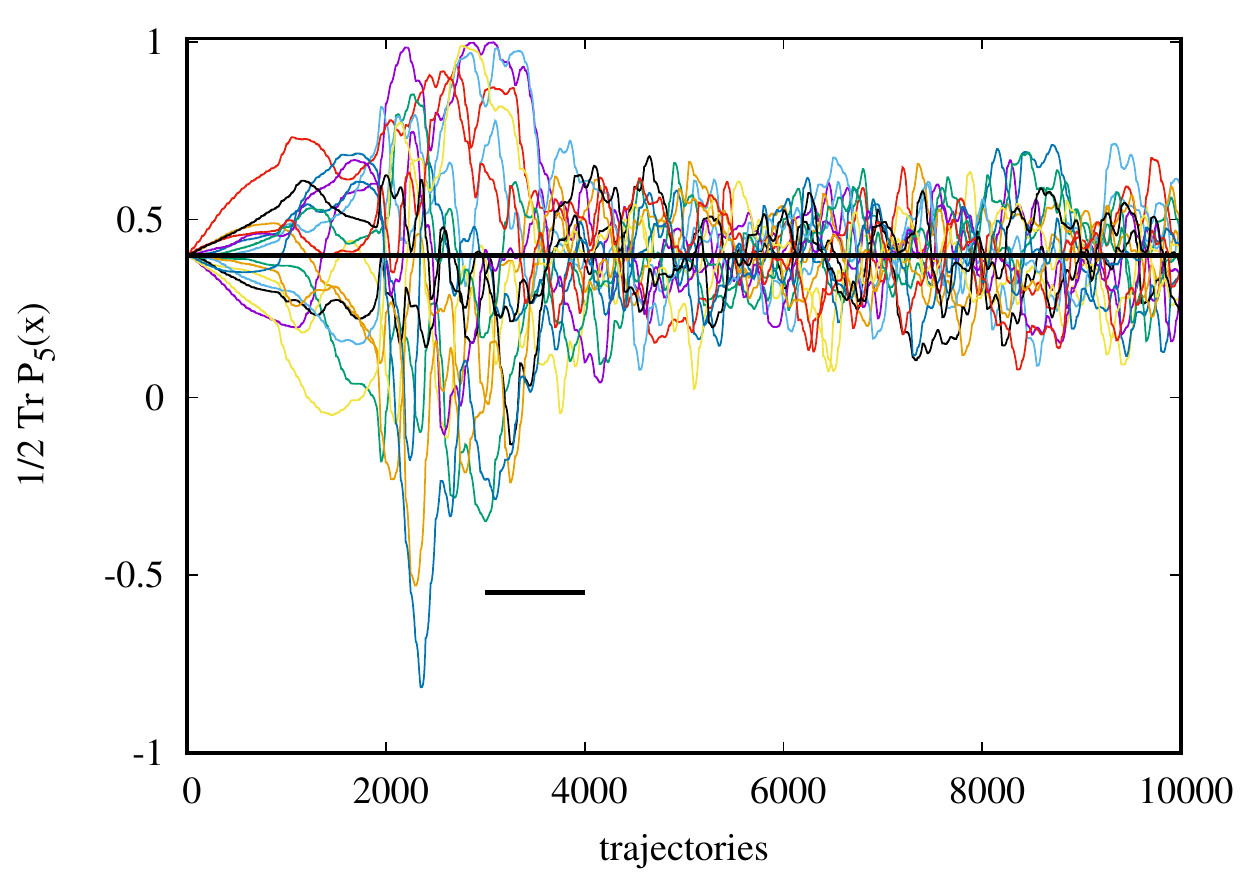}
 \put (47,21) {\scriptsize $\Phi=\dfrac{1}{2\Omega}\sum_{n_\mu}\tr P_5(n_\mu)=0.4$}
\end{overpic}
\includegraphics[width=.5\textwidth]{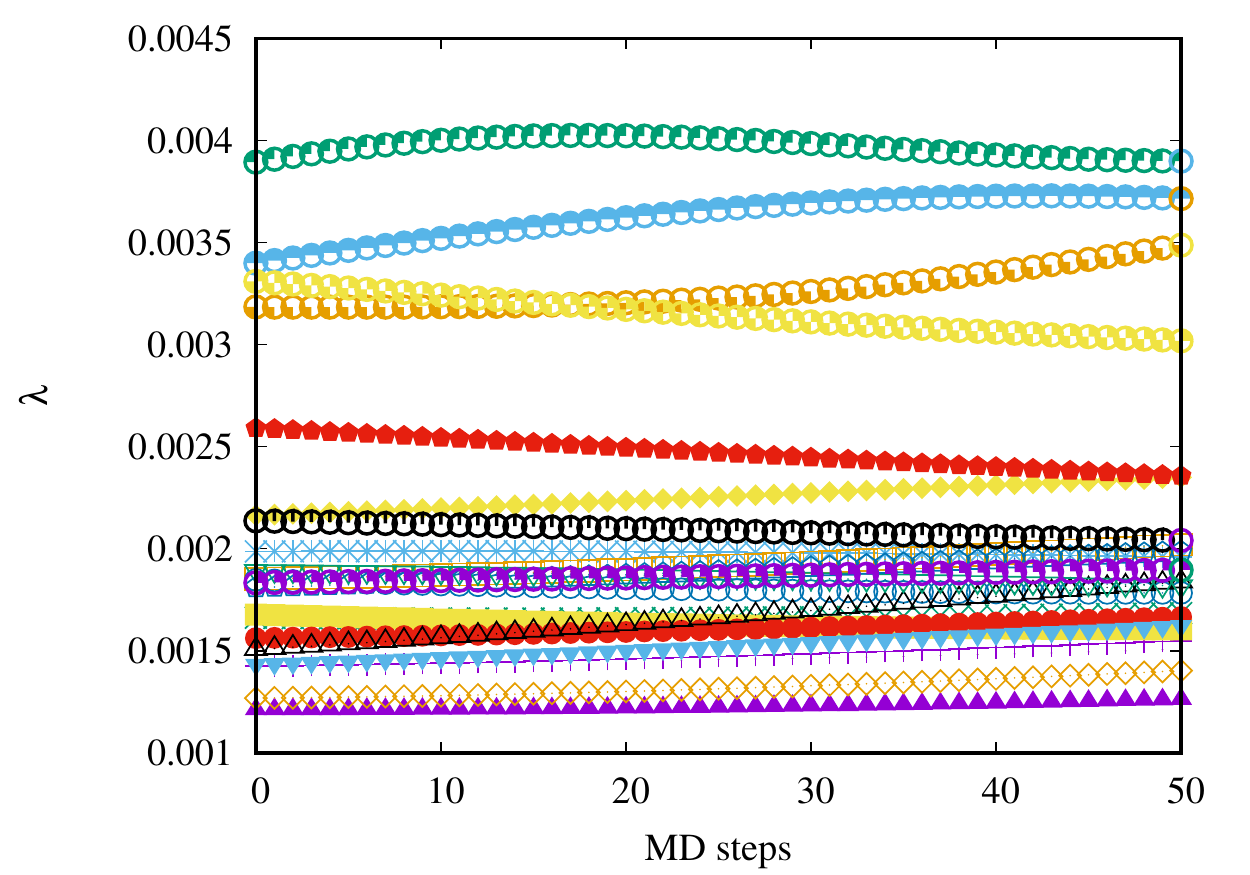}
\caption{5D Torus constrained HMC (Rattle) algorithm on a $2^5$ lattice: the individual Polyakov lines fluctuate around their average (left), guaranteed by an additional term in the Hamiltonian with the Lagrange multiplier $\lambda$ shown on the right evolving within different trajectories indicated by different colors/point styles.}\label{fig:rattlagr}
\end{figure}

Again, we have to initialize the Polyakov lines to fulfill the constraint condition (\ref{eq:trPcnst}), {\it e.g.}, with the help of axial gauge, and when drawing the Gaussian-distributed random conjugate momenta $\pi^r(n_\mu,n_5)$ we have to ensure that they comply with the hidden constraint (\ref{eq:torcnst2a}), which we achieve via orthogonal projection 
\bean
\pi_0(n_\mu,n_5)=\pi^r(n_\mu,n_5)-\dfrac{\sum_{n_\mu,n_5}\tr\{\tr[...\sigma_iU(n_\mu,n_5)...]\sigma^i\pi^r(n_\mu,n_5)\}}{\sum_{n_\mu,n_5}\tr\{(\tr[...\sigma_i U(n_\mu,n_5)...]\sigma^i)^2\}}\tr[...\sigma_i U(n_\mu,n_5)...]\sigma^i
\ean
\fig{fig:rattlagr} shows that the Rattle algorithm for the 5D torus keeps the average Polyakov loop 
fixed (left plot). The Lagrange multiplier along a trajectory is plotted on the right.

\begin{figure}[h]
\includegraphics[width=.5\linewidth]{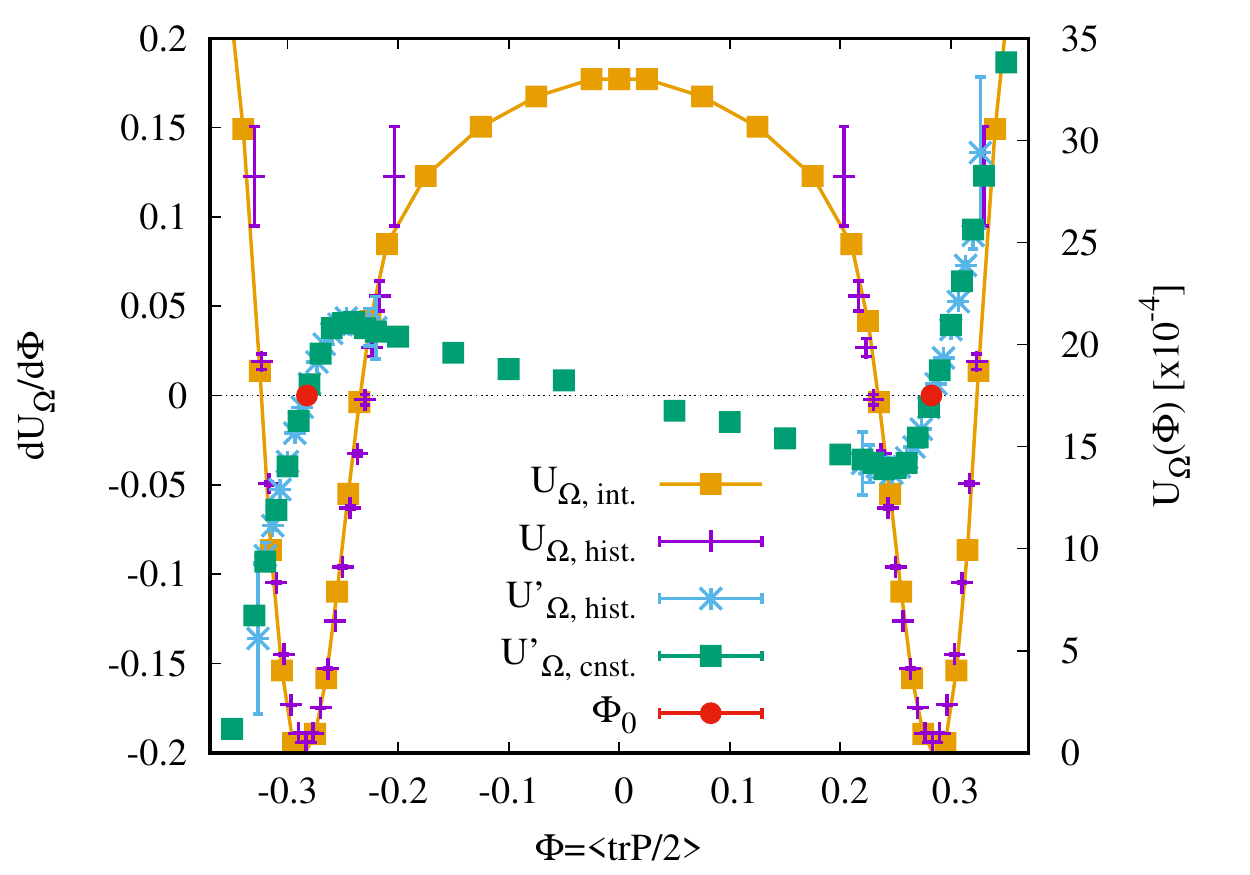}
\includegraphics[width=.5\linewidth]{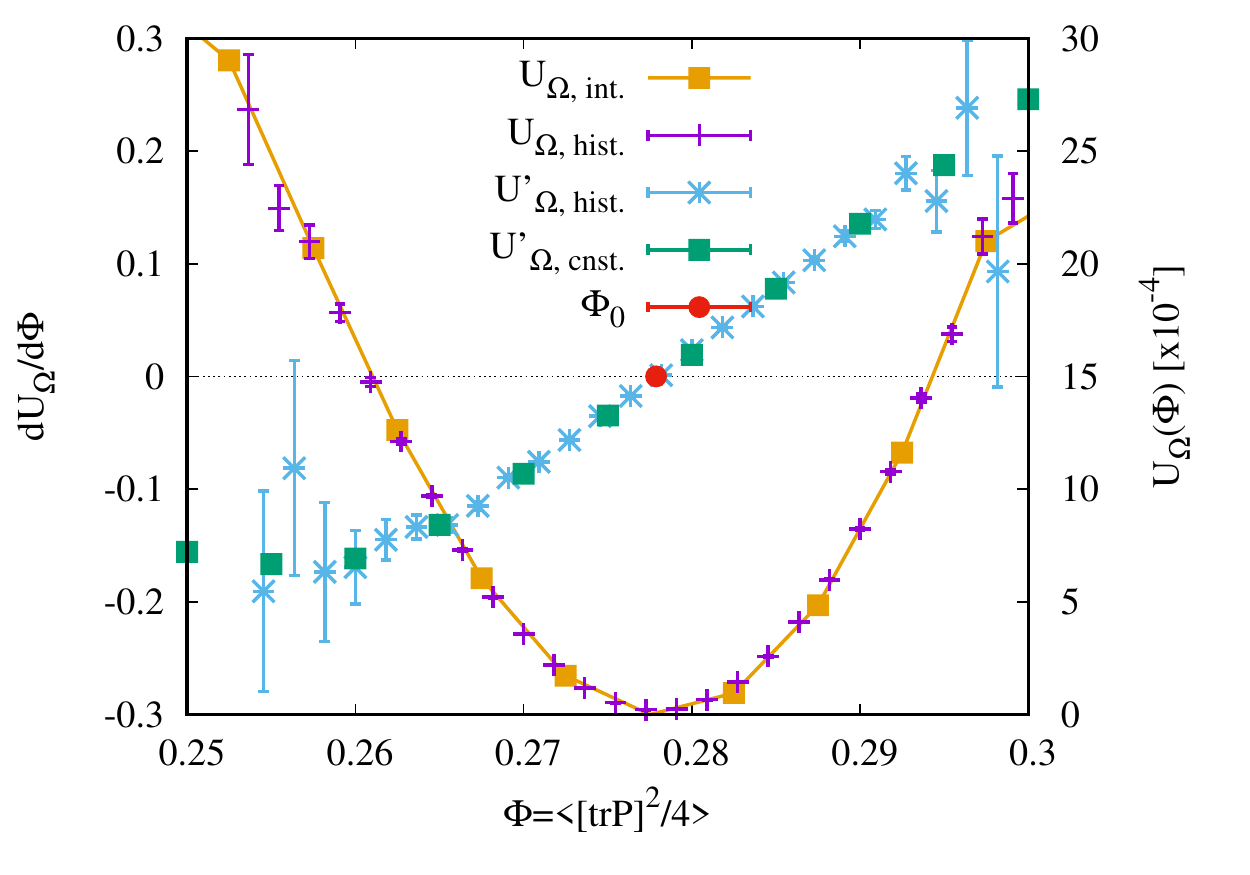}
\includegraphics[width=.5\linewidth]{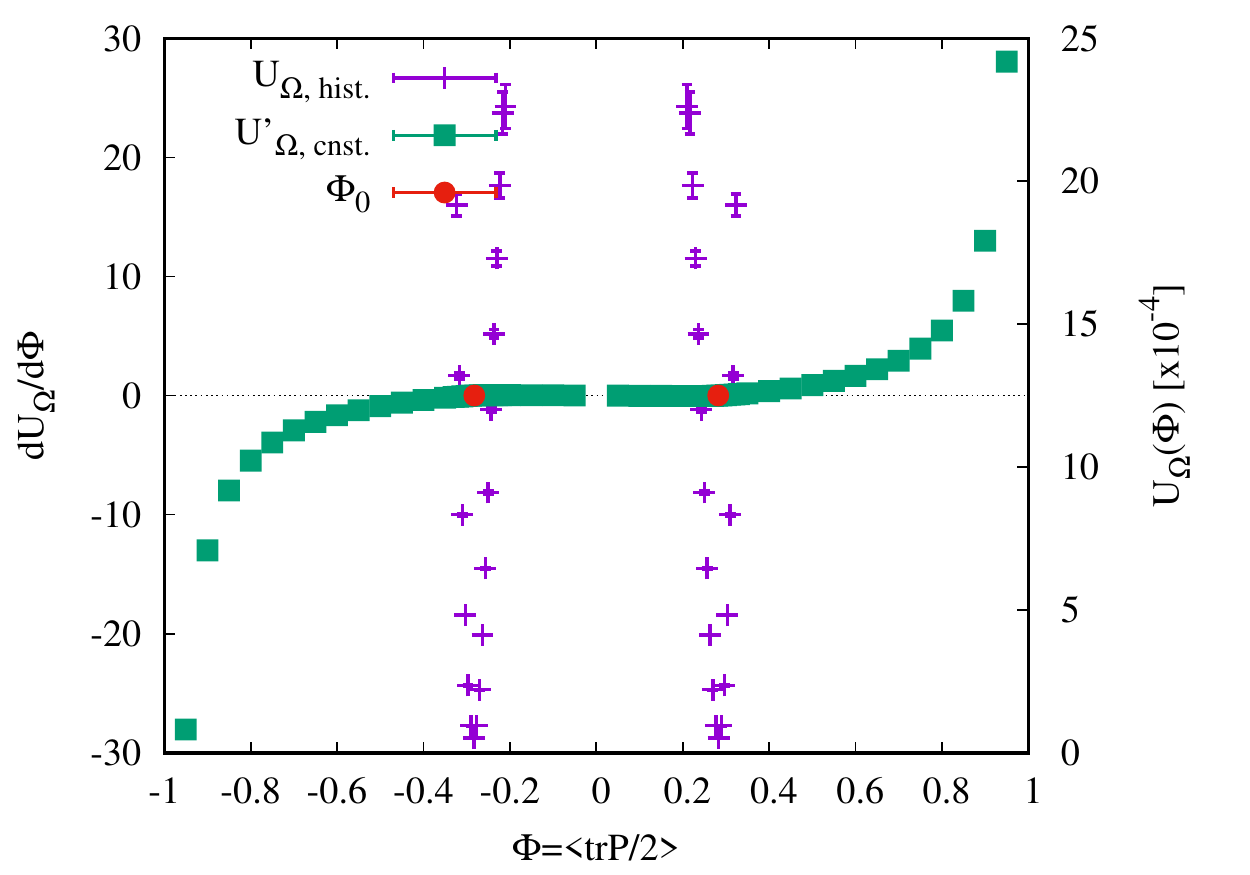}
\includegraphics[width=.5\linewidth]{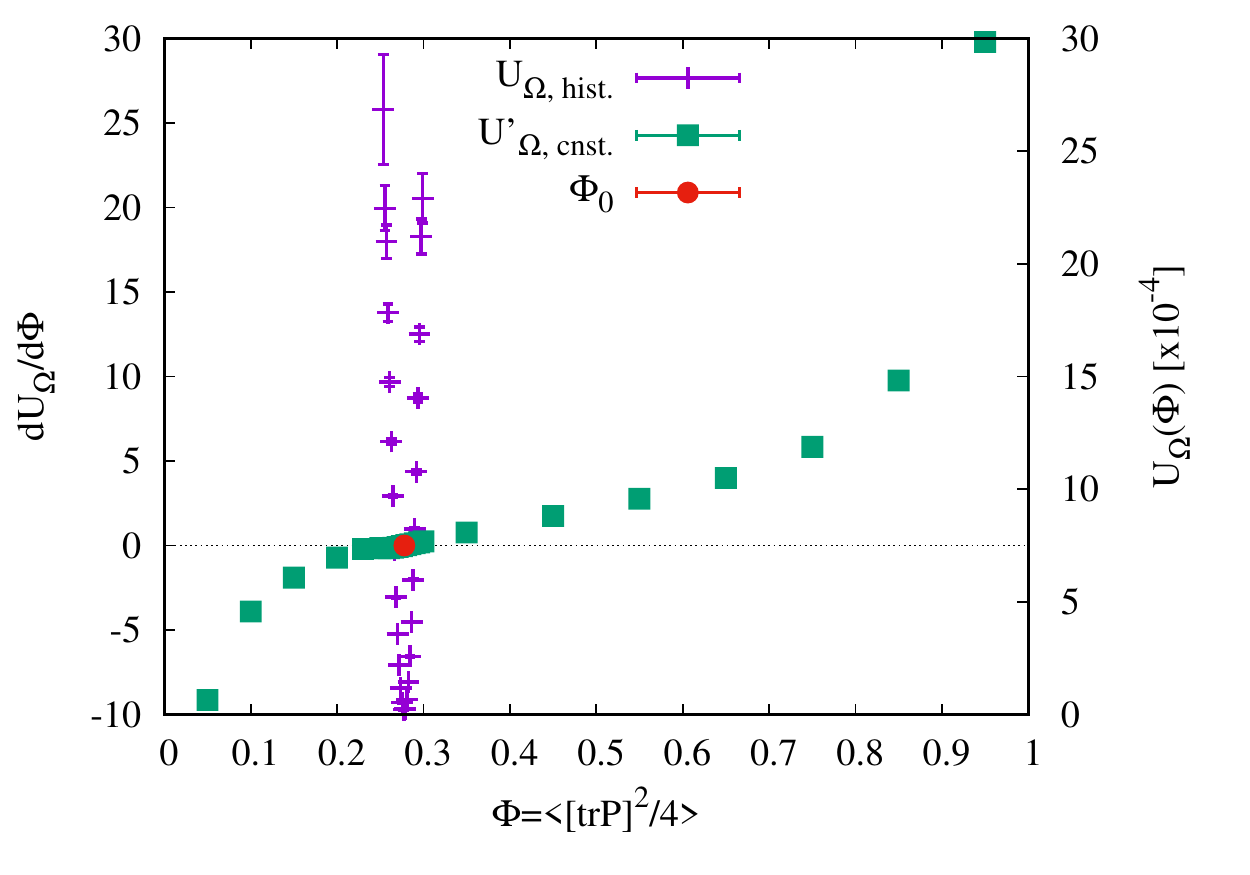}\\
\caption{5D Torus Effective Higgs Potential $U_{\Omega, \mathrm{hist.}}$ from the histogram method, its numerical derivative $U'_{\Omega, \mathrm{hist.}}$, the derivative from constrained HMC $U'_{\Omega, \mathrm{cnst.}}$ and its numerical integral $U_{\Omega, \mathrm{int.}}$ for $\beta_4=\beta_5=1.66, \Omega=8^4, N_5=4$. The unconstrained expectation value of the Polyakov $\Phi_0=\langle\tr P_5/2\rangle=\pm0.282$ has two degenerate minima (left column), while $\Phi_0=\langle(\tr P_5)^2/4\rangle=0.278$ is always greater than zero of course (right column).  
The potentials (and derivatives) diverge at the boundaries because $-1\leq\tr P_5/2\leq1$ and $0<(\tr P_5)^2/4\leq1$. 
The upper plots are a zoom of the plots in the second row.}\label{fig:b166}
\end{figure}

\begin{figure}[h]
\includegraphics[width=.5\linewidth]{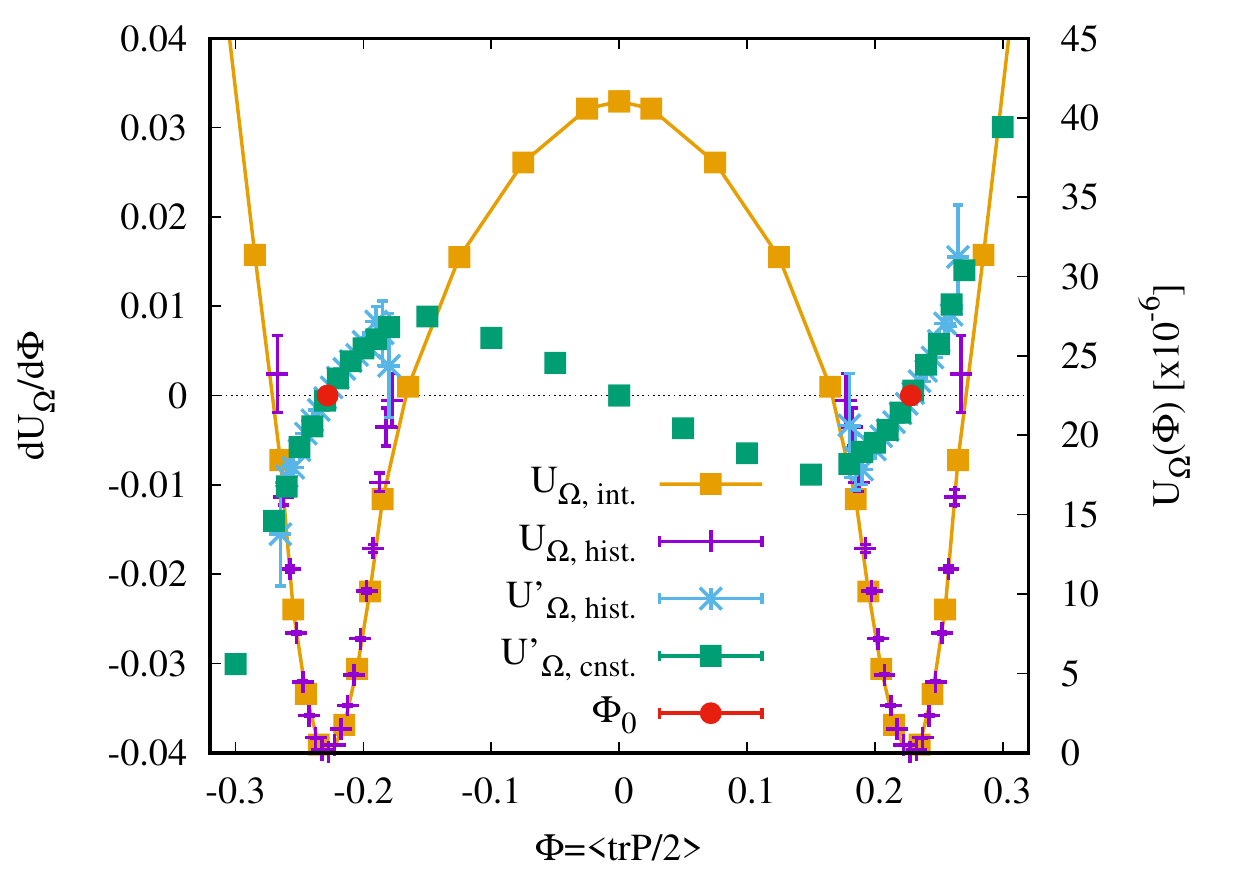}
\includegraphics[width=.5\linewidth]{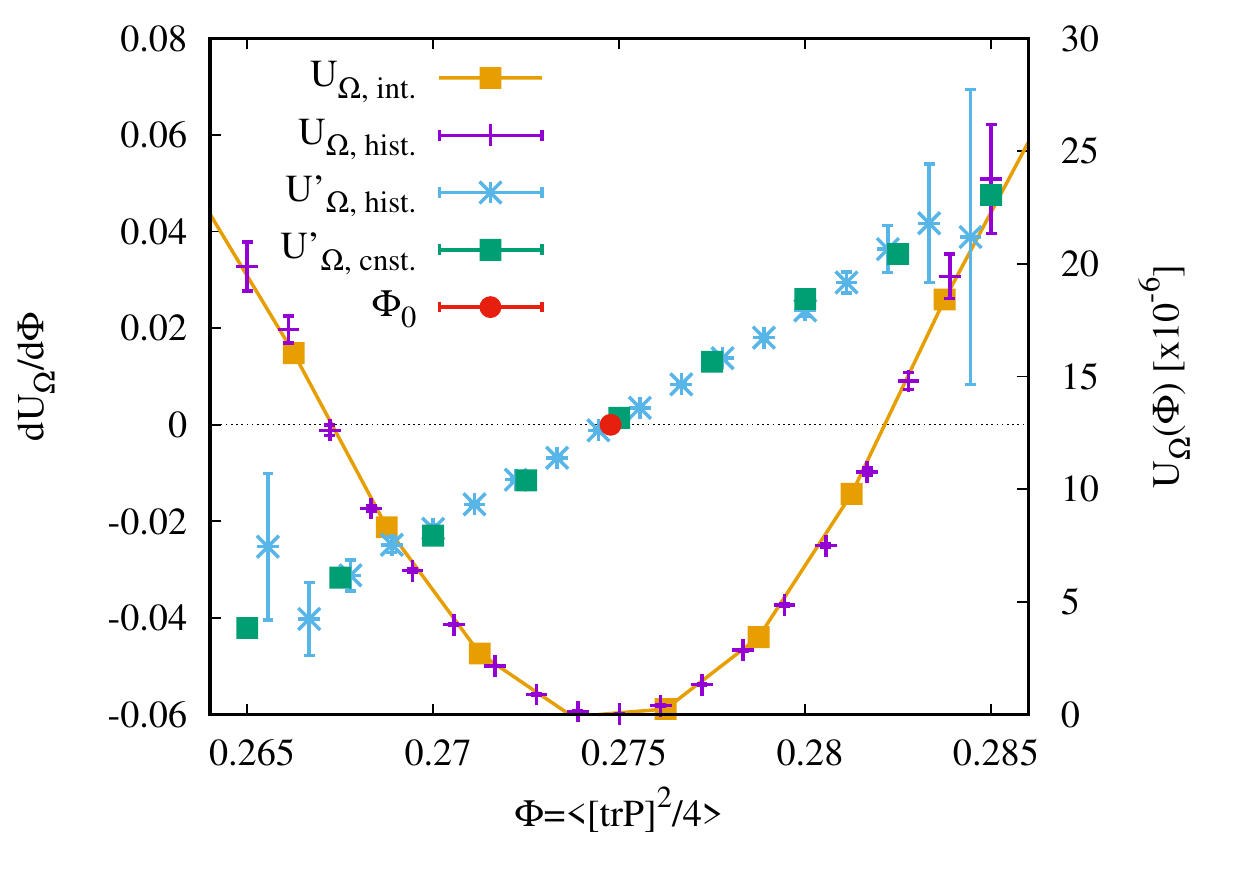}\\
\includegraphics[width=.5\linewidth]{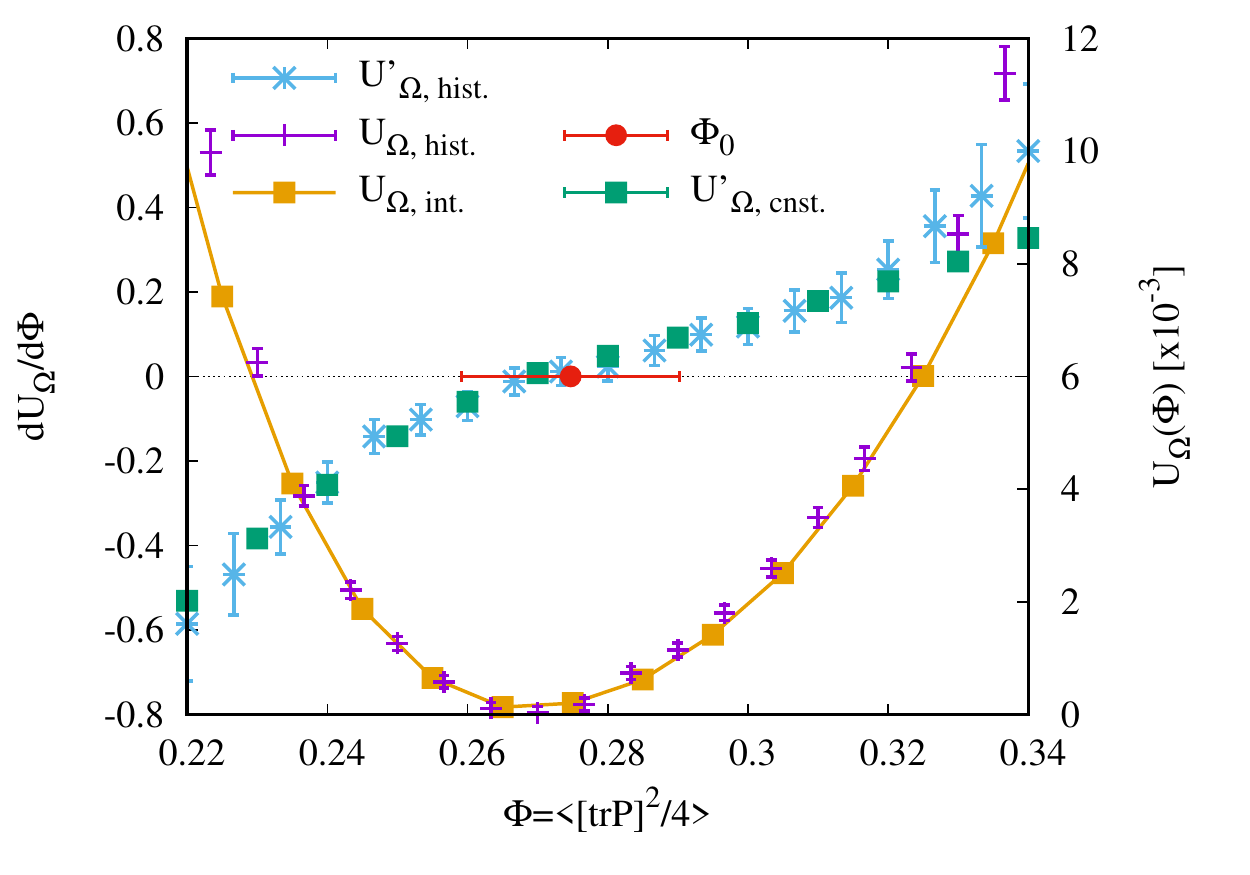}
\includegraphics[width=.5\linewidth]{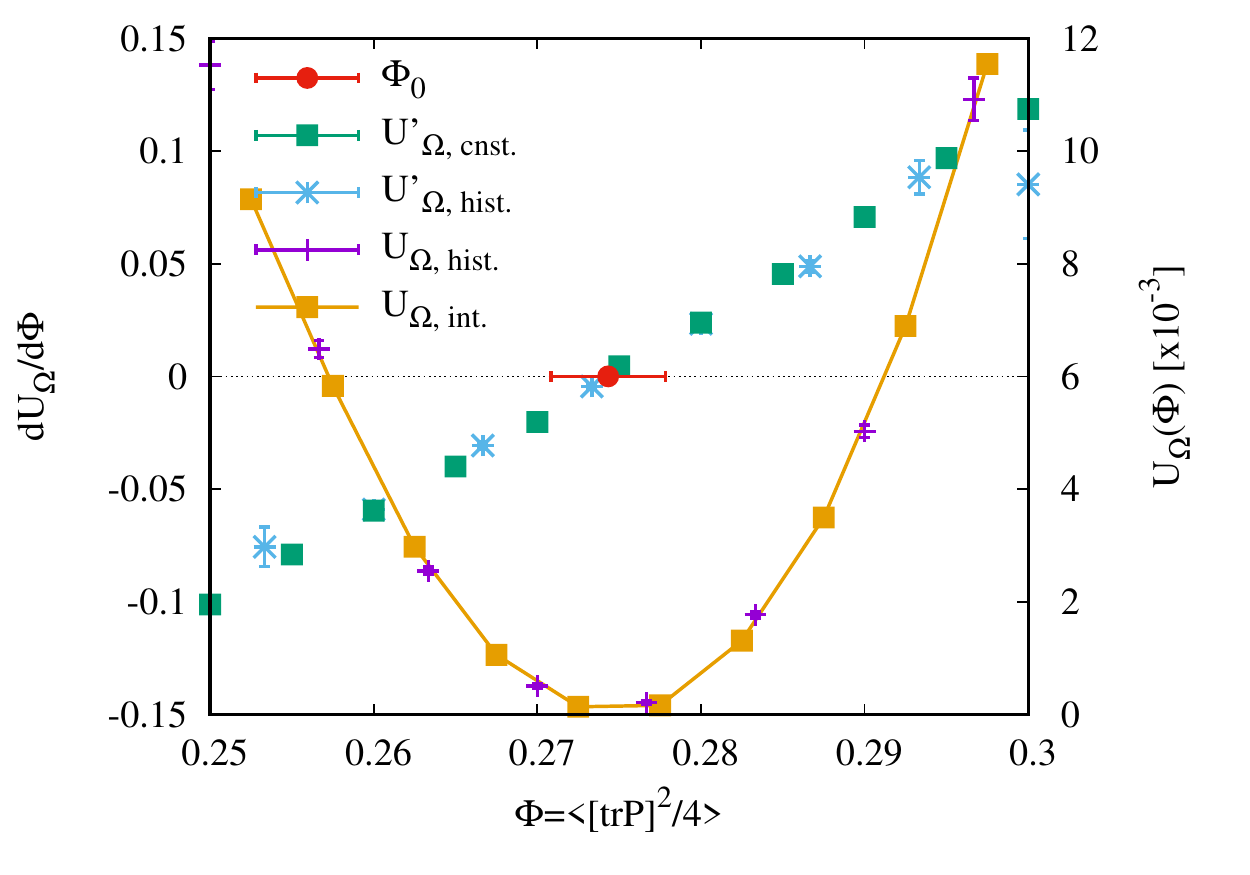}
\caption{5D Torus Effective Higgs Potential $U_{\Omega, \mathrm{hist.}}$ from the histogram method, its numerical derivative $U'_{\Omega, \mathrm{hist.}}$, the derivative from constrained HMC $U'_{\Omega, \mathrm{cnst.}}$ and its numerical integral $U_{\Omega, \mathrm{int.}}$ for $\beta_4=1.0, \beta_5=2.8, \Omega=24\times12^3, N_5=4$ (upper plots). This point lies in the compcat phase of the theory, where dimensional reduction via compactification is expected. For $\tr P$ we find two degenerate minima, $(\tr P)^2$ only shows the positive potential minimum, of course. The lower plots are on smaller volumes for comparison, {\it i.e.} $V=4^5$ on the left and $8^4\times4$ on the right. Note the different ranges for the abscissa and different scales for the ordinate.}\label{fig:b1b28}
\end{figure}

Because of center symmetry, $\langle\tr P_5\rangle$ always vanishes in finite volume. Therefore, we further investigate the constraint $\frac{1}{4\Omega}\sum_{n_\mu}[\tr P_5({n_\mu})]^2 =\Phi$ which fixes the Higgs field $\mathcal H=\frac{1}{8\Omega}\sum_{n_\mu}\tr[P_5({n_\mu})-P_5^\dagger({n_\mu})]^2=\Phi-1$ of the torus model and is invariant under center symmetry. The algorithm which fulfills the constraint is slightly more complicated than the one above and is formulated in appendix~\ref{app:constrP2}. The implementation is equivalent to the previous cases and, therefore, we just summarize the important steps:
\begin{itemize}
\item initialize the field variables $q$ to fulfill the constraint $g(q)=0$
\item draw (unconstrained) Gaussian distributed random momenta $p$
\item project the momenta $p$ to satisfy the hidden constraint $\dot g(q,p)$
\item propagate $p$ and $q$ as defined by the Rattle discretization
\item accept new fields with probability $r=\text{min}[1,\exp(-\Delta H)]$
\end{itemize}
With the right tools at hand we now measure the constraint effective potentials $U_\Omega(\Phi)$ via their derivatives $U'_\Omega(\Phi)=-\langle\lambda^{(1)}\rangle/\Omega$ for both cases $\langle\tr P_5/2\rangle$ and $\langle(\tr P_5)^2/4\rangle$ of the 5D SU(2) gauge theory on the torus.
In \fig{fig:b166} we show  effective Higgs potentials and their derivatives for the symmetric point $\beta_4=\beta_5=1.66$ on $\Omega=8^4, N_5=4$ lattices. The results of the constrained method are in good agreement with the histogram potential in the vicinity of the expectation value of the Higgs field. The histogram method is limited to that narrow region, getting narrower the larger the lattice size, while the constraint effective potential can be measured very precisely over the whole parameter range. 
The unconstrained expectation values of the Higgs field $\Phi_0$ exactly coincide with the zero crossing of the (constraint) effective potentials. $\Phi_0=\langle\tr P_5/2\rangle=\pm0.282$ is non-zero and we find two degenerate minima of the potential, the derivative of the constraint effective potential accordingly vanishes three times and its (numerical) integral shows the familiar Mexican hat form, which the histogram method cannot reproduce at all because of its limitations in potential width and accuracy.
The upper plots in \fig{fig:b1b28} present effective Higgs potentials for $\beta_4=1.0, \beta_5=2.8$ on $\Omega=24\times12^3, N_5=4$ lattices. This point lies in the so-called compact phase of the theory, where dimensional reduction via compactification is expected. Again we find two degenerate minima of the potential for $\Phi_0=\langle\tr P_5/2\rangle=\pm0.228$ and the Mexican hat form is reproduced by the constraint effective potential only. The Higgs observable $(\tr P)^2$ only shows the positive potential minimum, and we also show results for two smaller volumes for comparison (lower plots).
We conclude, that the constrained method is favorable to the histogram method in terms of accuracy and measurement range, the only drawback is the slower HMC algorithm which is essential for the first but not the latter.

\section{5D SU(2) GAUGE THEORY ON THE ORBIFOLD}\label{sec:chmcorb}

The orbifold theory we consider here is defined in the five-dimensional domain $I = \{n_\mu, 0 \leq n_5 \leq N_5\}$ with volume $N_t \times N_s^3 \times N_5$. The anisotropic Wilson gauge action for an SU(2) gauge theory on this orbifold is given by~\cite{Irges:2004gy}
\begin{equation}\label{aq:orbiaction}
S_W^{orb} = \sum_{n_\mu}\bigg[\frac{\beta_4}{2}\sum_{n_5=0}^{N_5}\sum_{\mu<\nu} w\,\mathrm{Re}\,\tr\{1-U_{\mu\nu}(n_\mu,n_5)\} +
\frac{\beta_5}{2}\sum_{n_5=0}^{N_5-1}\sum_{\mu}\mathrm{Re}\,\tr\{1-U_{\mu5}(n_\mu,n_5)\}\bigg] \,,
\end{equation}
which follows the parametrization of \eq{eq:toraction}. The weight $w$ is due to the orbifold geometry and takes a value $w=1/2$ for plaquettes $U_{\mu\nu}$ on the boundaries and it is $w=1$ elsewhere. The boundary links are in the gauge group U(1) and all other links are in SU(2). 
The anisotropy is $\gamma=\sqrt{\beta_5/\beta_4}$ and in the classical limit
$\gamma=a_4/a_5$, where $a_4$ denotes the lattice spacing in the usual
four dimensions and $a_5$ denotes the lattice spacing in the extra dimension.
The theory is defined on the interval $I=\{n_\mu,0\le n_5\le N_5\}$, where
$(n_\mu,n_5)$, $\mu=0,1,2,3$ are the integer coordinates of the points. 
Given a constrained Hamiltonian for the Polyakov loop 
\bea
P_5(n_\mu)&=&\prod_{n_5=0}^{N_5-1}[U_5(n_\mu,n_5)]\sigma_3\prod_{n_5=N_5-1}^{0}[U_5^\dagger(n_\mu,n_5)]\sigma_3\\
\tilde H[U_5]&=&S[U_5]+\sum_{n_\mu}\tr[\pi_5^2({\bf x})]+\lambda^{(1)}\bigg(\dfrac{1}{2\Omega}\sum_{n_\mu}\tr P_5(n_\mu) -\Phi\bigg)
\eea
we solve the constrained equations of motion
\bea
\dot U_5(n_\mu,n_5)&=&\pi_5(n_\mu,n_5)U_5(n_\mu,n_5),\nn
\dot\pi_5(n_\mu,n_5)&=&-\dfrac{\partial S[U_5]}{\partial U_5(n_\mu,n_5)}+\dfrac{\lambda^{(1)}}{8\Omega}\tr[...\sigma_iU_5(n_\mu,n_5)...-...U_5^\dagger(n_\mu,n_5)\sigma_i...]\sigma^i\non
\eea
using the Rattle algorithm derived in appendix~\ref{app:chmcorb}. Like in the torus models, we use a Secant method to determine the first Lagrange multiplier $\lambda^{(1)}$, starting with an educated guess given by \eq{aq:orbl1}, and we have to initialize the momenta according to the hidden constraint.

\begin{figure}[h!]
\includegraphics[width=.5\linewidth]{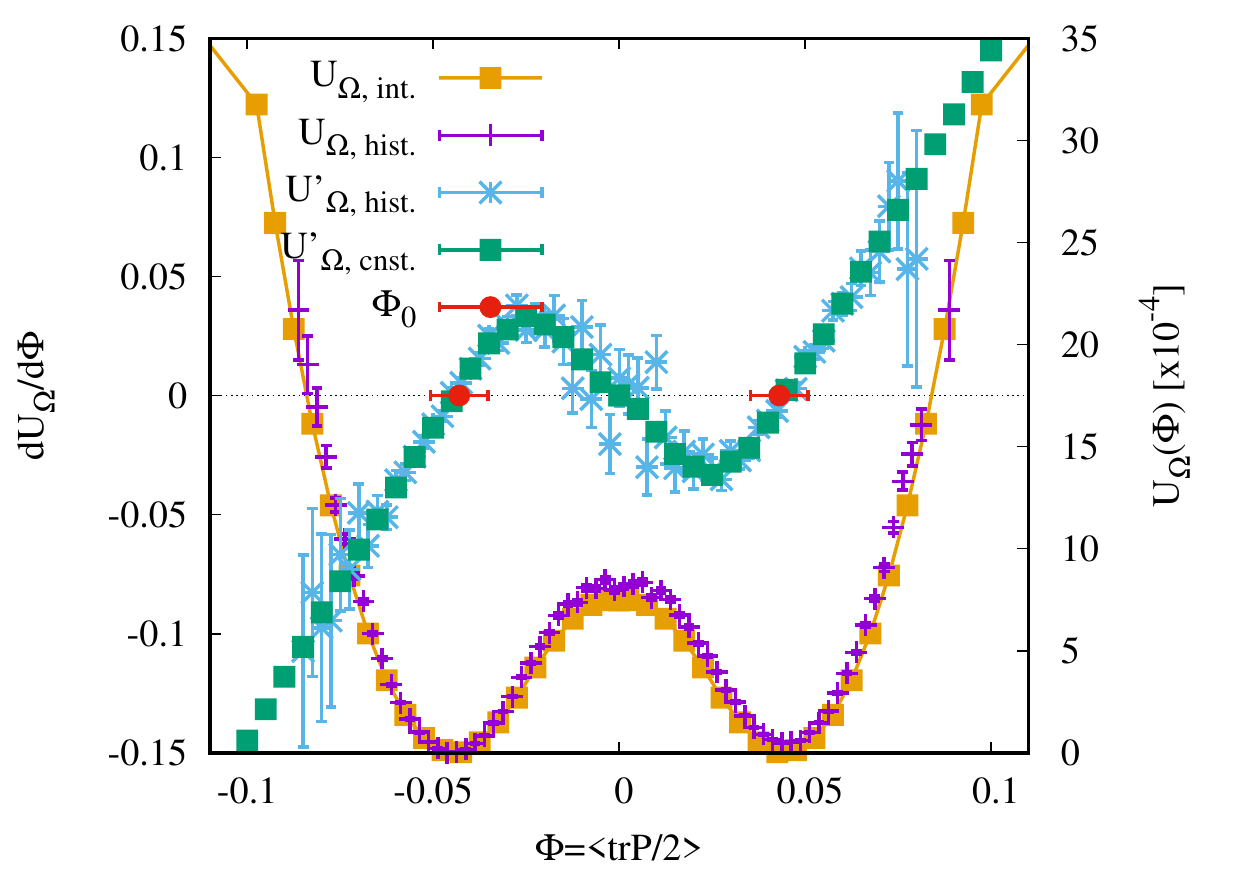}
\includegraphics[width=.5\linewidth]{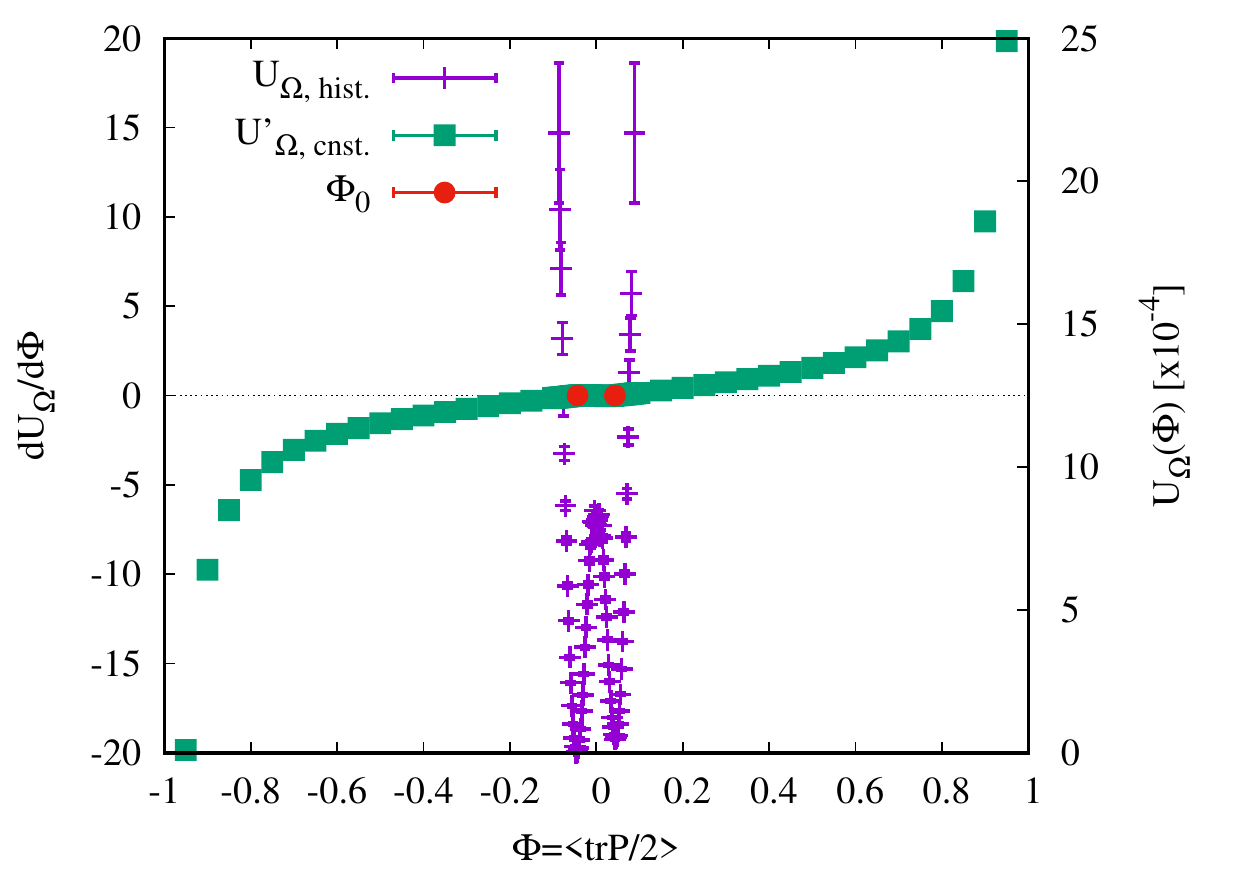}
\caption{5D Orbifold Effective Higgs Potential $U_{\Omega, \mathrm{hist.}}$ from the histogram method, its numerical derivative $U'_{\Omega, \mathrm{hist.}}$ and the derivative from constrained HMC $U'_{\Omega, \mathrm{cnst.}}=-\langle\lambda_n^{(1)}\rangle_\Phi/\Omega$ for the symmetric point $\beta_4=\beta_5=1.66, \Omega=8^4, N_5=5$. This point lies in the Higgs phase, where the stick symmetry is broken and we find two degenerate minima of the potential. The left plot is a zoom of the right plot, where we show the full parameter range $-1<\tr P/2<1$.}\label{fig:orb166}
\end{figure}

The constraint effective potential is measured via its derivative $U'_{\Omega, \mathrm{cnst.}}=-\langle\lambda_n^{(1)}\rangle_\Phi/\Omega$, a first result is shown in \fig{fig:orb166} for the symmetric point $\beta_4=\beta_5=1.66$ on a $\Omega=8^4, N_5=4$ lattice. This point in parameter space lies in the Higgs phase close to the bulk-driven phase transition, see \fig{fig:pds}. The stick symmetry is broken and we find two degenerate minima of the potential. 
The constrained observable $\Phi$ however, is not yet the exact definition of the Higgs field in the orbifold model, which is given by $\mathcal H=\frac{1}{4\Omega}\sum_{n_\mu}\tr[P_5({n_\mu})-P_5^\dagger({n_\mu}),\sigma_3]^2$. The algorithm for the latter is in progress.

\newpage

\section{CONCLUSIONS AND OUTLOOK}\label{sec:chmcconcl}

We successfully implemented constrained hybrid Monte Carlo algorithms for the 4D Abelian-Higgs and a 5D SU(2) gauge theory with torus and orbifold boundary conditions which allow us to simulate systems with constraint conditions, which appear as Lagrange multiplier terms in the Hamiltonian. To our knowledge, this is the first time this problem has been solved for theories with gauge fields. In order to solve the constrained equations of motion we use an extension of the Newton-St\"ormer-Verlet-leapfrog method to general Hamiltonians for constrained systems, the so-called Rattle algorithm.  
This generalized leap-frog method has an additional half integration step for the conjugate momenta in order to evaluate the so-called hidden constraint, which is the derivative of the constraint condition with respect to molecular dynamics time and in our cases involves both fields and momenta, after a full integration time step, in order to calculate the Lagrange multipliers which ensure that the constraints are fulfilled. The algorithm fulfills all necessary geometrical properties, summarized in appendix~\ref{app:rattle}, and we numerically tested the time-reversibility and volume preservation. First simulation results show that the constraint effective potential accurately not only reproduces the effective potential from the histogram method, but drastically increases the range of accessibility of the effective potential, as the histogram method is restricted to the vicinity of the expectation value of the Higgs field, and the precision does not deteriorate when increasing the volume. Furthermore, for the 4D Abelian-Higgs model in unitary gauge we show that the constraint effective potential agrees well with the continuum one-loop Higgs potential given in \eq{eq:1loopV} from~\cite{Irges:2017ztc}. We quantitatively compare the shape of the potentials for a weak gauge coupling. The Higgs mass determined from the potential agrees with the one extracted from fitting the two-point function of Higgs operators. 
The comparison to perturbative results is non-trivial for the other models due to the composite nature of the Higgs observables and, therefore, similar investigations are postponed to future work. In particular, we want to compare our constraint effective potentials in the 5D gauge theory cases with the one-loop effective Higgs potentials for the torus~\cite{deForcrand:2010be} and orbifold~\cite{Irges:2006hg} models respectively.

We also plan to measure the constraint effective potentials on larger lattices and extract the Higgs masses in the different models considered. The latter is given by the second derivative of the constraint effective potential at the vacuum expectation value of the Higgs field and can be compared to the masses measured by different methods, {\it e.g.}, by fitting two point functions from unconstrained simulations, which will allow us to study renormalization effects of the different mass determinations. Further we can compare the potentials measured in the different models and study the compactification and dimensional reduction scenarios of the 5D torus and orbifold models respectively via their connection to the 4D adjoint and Abelian-Higgs model. 

Finally, an interesting application of these algorithms are effective Polyakov loop actions in finite temperature QCD. The effective Polyakov loop action (PLA) is the theory which results from integrating out all of the degrees of freedom of the theory, subject to the condition that the Polyakov lines are held fixed. This was studied in the strong coupling expansion \cite{Bergner:2015rza}, but also in full lattice QCD simulations. It was found that this effective theory is more tractable than the underlying lattice gauge theory (LGT) when confronting the sign problem at finite density, for recent advances see~\cite{Greensite:2017qfl}.
The developed algorithms in this article can be adapted to this problem, where the individual Polyakov lines and not their average over the whole lattice are constrained. This requires Lagrange multiplier terms for each Polyakov line, which appear as a product in the path integral or a sum in the Hamiltonian. There is no additional numerical effort though, instead of summing over the whole lattice to evaluate one Lagrange multiplier, one just calculates the individual factors locally. The extraction of the effective Polyakov loop potential can in principal proceed in a similar way as described in this work, or else by the relative weights approach as presented in~\cite{Hollwieser:2016hne}.

\section{ACKNOWLEDGMENTS}
We thank Nikos Irges, Jean Zinn-Justin, Tomasz Korzec, Julius Kuti and Andreas Wipf for helpful discussions. We gratefully acknowledge the Gauss Center for Supercomputing (GCS) for providing computer time at the supercomputers JURECA/JUWELS at the Juelich Supercomputing Centre (JSC) under GCS/NIC project ID HWU24. This work was supported by the Deutsche Forschungsgemeinschaft in the SFB/TRR55 under Project B5 (R.H.).

\newpage

\begin{appendices}

\numberwithin{equation}{subsection}

\section{The Rattle algorithm for general constrained Hamiltonian systems}\label{app:rattle}

The most important numerical method for the solution of constrained Hamiltonian systems, the Rattle algorithm, is an adaptation of the Newton-St\"ormer-Verlet-leapfrog method that can be interpreted as a partitioned Runge-Kutta method and thus allows the extension to general Hamiltonians, c.f.~\cite{hairer:2002gni,hairer:2003gni}. 
We consider mechanical systems with coordinates $q$ that are subject to constraints $g(q) = 0$, and corresponding momenta $p$. The equations of motion are then given by
\bea
\dot p&=&-\nabla_qH(p,q)-\nabla_qg(q)\lambda\nn
\dot q&=&\nabla_pH(p,q),\qquad0=g(q),\label{aq:genEOMs}
\eea
where the Hamiltonian $H(p,q)$ is of the form 
\bea
H(p,q)=\frac{1}{2}p^TM^{-1}p+U(q)\label{aq:genH}
\eea
with a positive definite mass matrix $M$ and a potential $U(q)$. 
To compute the Lagrange multiplier $\lambda$, we differentiate the constraint 
$g(q(t))$ with respect to time, giving the so-called hidden constraint
\bea
0=\nabla_qg(q)^T\nabla_pH(p,q),\label{aq:hidden}
\eea
which is an invariant of the flow (\ref{aq:genEOMs}). We choose a step size h and discretized integration time $t_n=t_0+nh$. For initial values $(p_n, q_n)\in\mathcal M$, {\it i.e.}, consistent with $g(q)=0$ and (\ref{aq:hidden}), the Rattle method yields an approximation $(p_{n+1}, q_{n+1})$ which is again on the solution manifold $\mathcal M$:
\bsa
p_{n+1/2}&=&p_n+\dfrac{h}{2}\bigg(\nabla_qU(q_n)+\nabla_qg(q_n)\lambda_n^{(1)}\bigg)\\
q_{n+1}&=&q_n+hM^{-1}p_{n+1/2}, \\
\label{rattle.constraint1}
 0 & = &  g(q_{n+1}),\\
p_{n+1} &=& p_{n+1/2} + \frac{h}{2} \bigg( \nabla_q U(q_{n+1}), 
+\nabla_qg(q_{n+1})\lambda_n^{(2)}\bigg)\\ \label{rattle.constraint2}
0&=&\nabla_qg(q_{n+1})^T M^{-1} p_{n+1}
\esa
The first three equations determine $(p_{n+1/2},q_{n+1},\lambda_n^{(1)})$, whereas the remaining two give $(p_{n+1},\lambda_n^{(2)})$. Note that both Lagrangian multipliers are only intermediate variables and are not transported in the flow $\Phi_h$. We thus have a numerical flow $\Phi_h:\mathcal M\rightarrow\mathcal M$ with the following geometrical properties:
\begin{itemize}
\item the Rattle method is time-reversible, i.e., it holds $\rho \circ \Phi_h \circ \rho \circ \Phi_h= I $ with flipping the momenta denoted by $\rho(p,q)=(-p,q)$ ; 
if symmetry holds, {\it i.e.}, $\Phi_h=\Phi_{-h}^{-1}$, this is equivalent to 
$\rho \circ \Phi_h = \Phi_{-h} \circ \rho$; 
\begin{itemize}
    \item symmetry can be checked by exchanging the subscripts $n \leftrightarrow n+1$ and step size $h \leftrightarrow -h$, which has to leave the method unaltered. In our case, the first equation becomes the forth and vice-versa, if we change  the denomination of both Lagrangian multipliers, which are only intermediate variables; the second equation remains unchanged; the nonlinear equations~\eqref{rattle.constraint1},\eqref{rattle.constraint2} at time point $t_n+1$
    are become those at $t_n$ (and vice-versa).
    \item the condition $\rho \circ \Phi_h = \Phi_{-h} \circ \rho$ can be easily checked.
\end{itemize}
\item ensures long-time energy conservation, to be verified via $\langle\exp(-\Delta H)=1\rangle$
\item the Rattle algorithm is symplectic, {\it i.e.}, the flow preserves areas in phase space
\item det $\partial\Phi_h/\partial(p,q)=1$, {\it i.e.}, the flow preserves the volume in phase space
\item conservation of first integrals and preservation of adiabatic invariants
\item provides a discrete virial theorem
\end{itemize}
For more details see~\cite{hairer:2002gni,hairer:2003gni}. Now we summarize the algorithms for the various models\footnote{The algorithms for the 5D gauge theories presented in the proceedings~\cite{Hollwieser:2018nrh} differ from the present ones since they were the axial gauge.}.

\subsection{Rattle algorithm for the 4D Abelian-Higgs model}\label{app:abel}
\medskip
Given a constraint condition for the complex variables $\phi(n_\mu)$
\bea
\dfrac{1}{\Omega}\sum_{n_\mu}\phi^\dagger(n_\mu)\phi(n_\mu)=\Phi,
\eea 
the constrained HMC (Rattle) algorithm can be formulated in the following way
\bsa
\pi_{n+1/2}&=&\pi_n-\dfrac{h}{2}\bigg(\dfrac{\partial S}{\partial\phi_n}+\dfrac{2\phi_n\lambda_n^{(1)}}{\Omega}\bigg)\\
\phi_{n+1}&=&\phi_n+h\pi_{n+1/2}\\
0&=&\dfrac{1}{\Omega}\sum_{n_\mu}\phi_{n+1}^\dagger\phi_{n+1} - \Phi \\
\pi_{n+1}&=&\pi_{n+1/2}-\dfrac{h}{2}\bigg(\dfrac{\partial S}{\partial\phi_{n+1}}+\dfrac{2\phi_{n+1}\lambda_n^{(2)}}{\Omega}\bigg)\\
0&=&\dfrac{2}{\Omega}\sum_{n_\mu}\phi_{n+1}\pi_{n+1}\label{aq:hido}
\label{aq:rattleAH}
\esa
Plugging $\pi_{n+1/2}$ into $\phi_{n+1}$ and evaluating the constraint gives $\lambda_n^{(1)}$:
\bean
0&=&\sum\bigg(\dfrac{\phi_n}{h}+\pi_n-\dfrac{h}{2}\dfrac{\partial S}{\partial\phi_n}-\dfrac{h\phi_n\lambda_n^{(1)}}{\Omega}\bigg)^2-\dfrac{\Omega \Phi}{h^2}\\
&=&\cancel{\sum\dfrac{\phi_n^2}{h^2}}+\sum\pi_n^2+\sum\dfrac{h^2}{4}\bigg(\dfrac{\partial S}{\partial\phi_n}\bigg)^2+\dfrac{h^2\lambda_n^{(1)2}}{\Omega}\underbrace{\sum\dfrac{\phi_n^2}{\Omega}}_{=\Phi}+\dfrac{2}{h}\underbrace{\sum\phi_n\pi_n}_{=0}-\sum\phi_n\dfrac{\partial S}{\partial\phi_n}\\
&&-2\lambda_n^{(1)}\underbrace{\sum\dfrac{\phi_n^2}{\Omega}}_{=\Phi}-h\sum\pi_n\dfrac{\partial S}{\partial\phi_n}-\dfrac{2h\lambda_n^{(1)}}{\Omega}\underbrace{\sum\phi_n\pi_n}_{=0}+\dfrac{h^2\lambda_n^{(1)}}{\Omega}\sum\phi_n\dfrac{\partial S}{\partial\phi_n}-\cancel{\dfrac{\Omega \Phi}{h^2}}\\
&=&\lambda_n^{(1)2}+\lambda_n^{(1)}\bigg(\sum\dfrac{\phi_n}{\Phi}\dfrac{\partial S}{\partial\phi_n}-\dfrac{2\Omega}{h^2}\bigg)
+\dfrac{\Omega}{\Phi}\sum\bigg(\dfrac{\pi_n^2}{h^2}-\dfrac{\phi_n}{h^2}\dfrac{\partial S}{\partial\phi_n}-\dfrac{\pi_n}{h}\dfrac{\partial S}{\partial\phi_n}+\dfrac{1}{4}\bigg(\dfrac{\partial S}{\partial\phi_n}\bigg)^2\bigg)
\ean
\bean
\Rightarrow\lambda_n^{(1)}&=&\dfrac{\Omega}{h^2}-\sum\dfrac{\phi_n}{2\Phi}\dfrac{\partial S}{\partial\phi_n}\pm\sqrt{\dfrac{\Omega^2}{h^4}+\bigg(\sum\dfrac{\phi_n}{2\Phi}\dfrac{\partial S}{\partial\phi_n}\bigg)^2-\dfrac{\Omega}{\Phi}\sum\bigg(\dfrac{\pi_n}{h}-\dfrac{1}{2}\dfrac{\partial S}{\partial\phi_n}\bigg)^2}
\ean
During numerical simulations it turns out that only the $-$ sign in front of the square root in $\lambda_n^{(1)}$ fulfills the constraint condition. Plugging $\pi_{n+1}$ into the hidden constraint (\ref{aq:hido}) gives
\bean
\lambda_n^{(2)}&=&\sum_{n_\mu}\bigg(\dfrac{\phi_{n+1}\pi_{n+1/2}}{h\Phi}-\dfrac{\phi_{n+1}}{2\Phi}\dfrac{\partial S}{\partial\phi_{n+1}}\bigg)
\ean

When drawing the Gaussian-distributed random conjugate momenta $\pi^r(n_\mu)$ we have to ensure that they comply with the hidden constraint, which we achieve via orthogonal projection
\bean
\pi_0(n_\mu)=\pi^r(n_\mu)-\dfrac{\phi(n_\mu)}{\Omega\Phi}\sum_{m_\mu}\pi^r(m_\mu)\phi(m_\mu),
\ean 
as can be verified by plugging it back into (\ref{aq:hido}).

\subsection{Rattle algorithm for 5D SU(2) gauge theory on the torus fixing $\langle\tr P\rangle$}\label{app:chmctor}
\medskip
Given a constraint condition for the SU(2) link variables $U_5(n_\mu)$ in the fifth dimension
\bea
\dfrac{1}{2\Omega}\sum_{n_\mu}\tr\prod_{n_5=0}^{N_5-1} U_5(n_\mu,n_5)=\Phi\label{aq:torcnst}
\eea
the constrained HMC (Rattle) algorithm can be formulated in the following way
\bsa
\pi_{n+1/2}&=&\pi_n-\dfrac{h}{2}\bigg(\dfrac{\partial S}{\partial U_n}-\dfrac{\lambda_n^{(1)}}{8\Omega}\tr[...\sigma_i U_n...]\sigma^i\bigg)\label{aq:momupdtor}\\ 
U_{n+1}&=&e^{h\pi_{n+1/2}}U_n\label{aq:expupdtor}\\
0 &=& \dfrac{1}{2\Omega}\sum_{n_\mu}\tr\prod_{n_5=0}^{N_5-1} U_{n+1}(n_\mu,n_5)-\Phi \\
\pi_{n+1}&=&\pi_{n+1/2}-\dfrac{h}{2}\bigg(\dfrac{\partial S}{\partial U_{n+1}}-\dfrac{\lambda_n^{(2)}}{8\Omega}\tr[...\sigma_i U_{n+1}...]\sigma^i\bigg)\label{aq:pin1}\\
0&=&\sum_{n_\mu,n_5}\tr\{\tr[...\sigma_i U_{n+1}(n_\mu,n_5)...]\sigma^i\pi_{n+1}(n_\mu,n_5)\} \label{aq:torcnst2}
\esa
where $\tr[...\sigma_iU_5(n_\mu,n_5)...]\sigma^i$ is the derivative of the constraint with respect to $U_5(n_\mu,n_5)$, using
\bean
\partial_5\tr U_5&\equiv&\dfrac{i}{2}\sigma^a\partial_{5,a}\tr U_5=\dfrac{i}{2}\sigma^a\lim_{\epsilon\rightarrow0}\dfrac{\tr(e^{i\epsilon\frac{\sigma_a}{2}}U_5)-\tr U_5}{\epsilon}
=-\dfrac{1}{4}\tr(\sigma_aU_5)\sigma^a
\ean
When evaluating the constraint condition for $U_{n+1}$ we truncate the exponential in (\ref{aq:expupdtor})
\bea
e^{h\pi_{n+1/2}}&=&1+h\pi_n-\dfrac{h^2}{2}\dfrac{\partial S}{\partial U_n}+\dfrac{\lambda_n^{(1)}h^2}{16\Omega}\tr[...\sigma_i U_n...]\sigma^i+\dfrac{h^2}{2}\pi_n^2+\mathcal{O}(h^3)\label{aq:exp}
\eea
and solve for $\lambda_n^{(1)}$ up to the same order 
\bea
\Phi&=&\dfrac{1}{2\Omega}\sum_{n_\mu}\tr[e^{h\pi_{n+1/2}(n_\mu,0)}U_n(n_\mu,0)\ldots e^{h\pi_{n+1/2}(n_\mu,N_5-1)}U_n(n_\mu,N_5-1)]\nn
\cancel{\Phi}&=&\dfrac{1}{2\Omega}\sum_{n_\mu}\bigg\{h\sum_{n_5=0}^{N_5-1}\bigg(\underbrace{\tr[...\pi_n(n_\mu,n_5)U_n(n_\mu,n_5)...]}_{\sum\ldots=0}-\dfrac{h}{2}\tr[...\dfrac{\partial S}{\partial U_n(n_\mu,n_5)}U_n(n_\mu,n_5)...]\nn
&&+\dfrac{\lambda_n^{(1)}h}{16\Omega}\tr\{...\tr[...\sigma_i U_n(n_\mu,n_5)...]\sigma^iU_n(n_\mu,n_5)...\}+\dfrac{h}{2}\tr[...\pi_n^2(n_\mu,n_5)U_n(n_\mu,n_5)...]\nn
&&+h\sum_{m_5>n_5}^{N_5-1}\tr[...\pi_n(n_\mu,n_5)U_n(n_\mu,n_5)...\pi_n(n_\mu,m_5)U_n(n_\mu,m_5)...]\bigg)+\cancel{\tr P_n(n_\mu)}\bigg\}+\mathcal{O}(h^3)\nn
\dfrac{\lambda_n^{(1)}}{8\Omega}&=&\bigg\{\sum_{n_\mu,n_5}\bigg(\tr[...\dfrac{\partial S}{\partial U_n(n_\mu,n_5)}U_n(n_\mu,n_5)...]-\tr[...\pi_n^2(n_\mu,n_5)U_n(n_\mu,n_5)...]\label{aq:lambda}\\
&&-2\sum_{m_5>n_5}^{N_5-1}\tr[...\pi_n(n_\mu,n_5)U_n(n_\mu,n_5)...\pi_n(n_\mu,m_5)U_n(n_\mu,m_5)...]\bigg)\bigg\}/\nn
&&\qquad\sum_{n_\mu,n_5}\tr\{...\tr[...\sigma_i U_n(n_\mu,n_5)...]\sigma^iU_n(n_\mu,n_5)...\}+\mathcal{O}(h^3)\non
\eea
We use a Secant method to get the Lagrange multiplier with machine precision. Plugging $\pi_{n+1}$ (\ref{aq:pin1}) into the hidden constraint (\ref{aq:torcnst2}) gives $\lambda_n^{(2)}$: 
\bea
\dfrac{\lambda_n^{(2)}}{8\Omega}&=&\sum_{n_\mu,n_5}\bigg(\tr[...\sigma_iU_{n+1}(n_\mu,n_5)...]\sigma^i\partial S/\partial U_{n+1}(n_\mu,n_5)\label{aq:mu}\\
&&\quad-2\tr[...\sigma_iU_{n+1}(n_\mu,n_5)...]\sigma^i\pi_{n+1/2}(n_\mu,n_5)/h\bigg)/\nn
&&\qquad\sum_{n_\mu,n_5}\tr\{(\tr[...\sigma_i U_{n+1}(n_\mu,n_5)...]\sigma^i)^2\}\non
\eea

Again, when drawing the Gaussian-distributed random conjugate momenta $\pi^r(n_\mu,n_5)$ we have to ensure that they comply with the hidden constraint, which we achieve by orthogonal projection 
\bean
\pi_0(n_\mu,n_5)&=&\pi^r(n_\mu,n_5)-\mu\tr[...\sigma_i U(n_\mu,n_5)...]\sigma^i
\ean
and solving for $\mu$ by plugging $\pi_0(n_\mu,n_5)$ into the hidden constraint (\ref{aq:torcnst2})
\bean
\mu&=&\dfrac{\sum_{n_\mu,n_5}\tr\{\tr[...\sigma_iU(n_\mu,n_5)...]\sigma^i\pi^r(n_\mu,n_5)\}}{\sum_{n_\mu,n_5}\tr\{(\tr[...\sigma_i U(n_\mu,n_5)...]\sigma^i)^2\}}
\ean

\medskip
\subsection{Rattle algorithm for 5D SU(2) gauge theory on the torus fixing $\langle(\tr P_5)^2\rangle$}\label{app:constrP2}
\medskip
Given a constraint condition for the average of the squared trace of the Polyakov loop $P_5$ in the extra dimension constructed from SU(2) link variables $U_5(n_\mu)$
\bea
\dfrac{1}{4\Omega}\sum_{n_\mu}(\tr P_{n+1})^2=\dfrac{1}{4\Omega}\sum_{n_\mu}\tr\prod_{n_5=0}^{N_5-1} U_{n+1}(n_\mu,n_5)\tr\prod_{n_5=0}^{N_5-1} U_{n+1}(n_\mu,n_5)=\Phi
\eea
the constrained HMC (Rattle) algorithm can be formulated in the following way
\bsa
\pi_{n+1/2}&=&\pi_n-\dfrac{h}{2}\bigg(\dfrac{\partial S}{\partial U_n}-\dfrac{\lambda_n^{(1)}}{16\Omega}\tr P_n\tr[...\sigma_i U_n...]\sigma^i\bigg)\\ 
U_{n+1}&=&e^{h\pi_{n+1/2}}U_n\label{aq:exptorupdt2}\\
0 &=& \dfrac{1}{4\Omega}\sum_{n_\mu}\tr\prod_{n_5=0}^{N_5-1} U_{n+1}(n_\mu,n_5)\tr\prod_{n_5=0}^{N_5-1} U_{n+1}(n_\mu,n_5)-\Phi \\
\pi_{n+1}&=&\pi_{n+1/2}-\dfrac{h}{2}\bigg(\dfrac{\partial S}{\partial U_{n+1}}-\dfrac{\lambda_n^{(2)}}{16\Omega}\tr P_{n+1}\tr[...\sigma_i U_{n+1}...]\sigma^i\bigg)\label{aq:pin2}\\
0&=&\sum_{n_\mu,n_5}\tr P_{n+1}\tr\{\tr[...\sigma_i U_{n+1}(n_\mu,n_5)...]\sigma^i\pi_{n+1}(n_\mu,n_5)\}\label{aq:cnstrP2}
\esa
with $2\tr P_5(n_\mu)\tr[...\sigma_iU_5(n_\mu,n_5)...]\sigma^i$ the derivative of the constraint with respect to $U_5(n_\mu,n_5)$.
Again, we truncate the exponential in (\ref{aq:exptorupdt2})
\bea
e^{h\pi_{n+1/2}}&=&1+h\pi_n-\dfrac{h^2}{2}\dfrac{\partial S}{\partial U_n}+\dfrac{\lambda_n^{(1)}h^2}{32\Omega}\tr P_n\tr[...\sigma_i U_n...]\sigma^i+\dfrac{h^2}{2}\pi_n^2+\mathcal{O}(h^3)\label{aq:exp}
\eea
and plug $U_{n+1}$ into the constraint condition, solving for $\lambda_n^{(1)}$ up to the same order 
\bean
\Phi&=&\dfrac{1}{4\Omega}\sum_{n_\mu}\bigg(\tr[e^{h\pi_{n+1/2}(n_\mu,0)}U_n(n_\mu,0)\ldots e^{h\pi_{n+1/2}(n_\mu,N_5-1)}U_n(n_\mu,N_5-1)]\bigg)^2\\
\cancel{\Phi}&=&\dfrac{1}{4\Omega}\sum_{n_\mu}\bigg\{\cancel{(\tr P_n)^2}+2h\sum_{n_5=0}^{N_5-1}\bigg(\underbrace{\tr[...\pi_nU_n...]}_{\sum\ldots=0}-\dfrac{h}{2}\tr[...\dfrac{\partial S}{\partial U_n(n_\mu,n_5)}U_n(n_\mu,n_5)...]\\
&&+\dfrac{\lambda_n^{(1)}h}{32}\tr P_n\tr\{...\tr[...\sigma_i U_n(n_\mu,n_5)...]\sigma^iU_n(n_\mu,n_5)...\}+\dfrac{h}{2}\tr[...\pi_n^2(n_\mu,n_5)U_n(n_\mu,n_5)...]\\
&&+h\sum_{m_5>n_5}^{N_5-1}\tr[...\pi_n(n_\mu,n_5)U_n(n_\mu,n_5)...\pi_n(n_\mu,m_5)U_n(n_\mu,m_5)...]\bigg)\,\tr P_n(n_\mu)\\
&&+h^2\sum_{n_5=0}^{N_5-1}\sum_{m_5=0}^{N_5-1}\tr[...\pi_n(n_\mu,n_5)U_n(n_\mu,n_5)...]\tr[...\pi_n(n_\mu,m_5)U_n(n_\mu,m_5)...]
\bigg\}+\mathcal{O}(h^3)\non
\ean
\bean
\dfrac{\lambda_n^{(1)}}{16\Omega}&=&\bigg\{\sum_{n_\mu}\bigg[\tr P_n(n_\mu)\sum_{n_5}\bigg(\tr[...\dfrac{\partial S}{\partial U_n(n_\mu,n_5)}U_n(n_\mu,n_5)...]-\tr[...\pi_n^2(n_\mu,n_5)U_n(n_\mu,n_5)...]\\
&&-2\sum_{m_5>n_5}^{N_5-1}\tr[...\pi_n(n_\mu,n_5)U_n(n_\mu,n_5)...\pi_n(n_\mu,m_5)U_n(n_\mu,m_5)...]\bigg)\\
&&-\sum_{n_5}\sum_{m_5}\tr[...\pi_n(n_\mu,n_5)U_n(n_\mu,n_5)...]\tr[...\pi_n(n_\mu,m_5)U_n(n_\mu,m_5)...]\bigg]\bigg\}\\&&\qquad/\sum_{n_\mu,n_5}(\tr P_n(n_\mu))^2\tr\{...\tr[...\sigma_i U_n(n_\mu,n_5)...]\sigma^iU_n(n_\mu,n_5)...\}+\mathcal{O}(h^3)\\
\ean

We use a Secant method to get the Lagrange multiplier with machine precision. Plugging $\pi_{n+1}$ (\ref{aq:pin2}) into the hidden constraint (\ref{aq:cnstrP2}) gives $\lambda_n^{(2)}$: 
\bean
\dfrac{\lambda_n^{(2)}}{16\Omega}&=&\sum_{n_\mu,n_5}\bigg(\tr P_n(n_\mu)\tr[...\sigma_iU_{n+1}(n_\mu,n_5)...]\sigma^i\partial S/\partial U_{n+1}(n_\mu,n_5)\\
&&\qquad-2\tr P_n(n_\mu)\tr[...\sigma_iU_{n+1}(n_\mu,n_5)...]\sigma^i\pi_{n+1/2}(n_\mu,n_5)/h\bigg)/\\
&&\qquad\sum_{n_\mu,n_5}(\tr P_n(n_\mu))^2\tr\{(\tr[...\sigma_i U_{n+1}(n_\mu,n_5)...]\sigma^i)^2\}
\ean

Again, when drawing the Gaussian-distributed random conjugate momenta $\pi^r(n_\mu,n_5)$ we have to ensure that they comply with the hidden constraint, which we achieve by the orthogonal projection 
\bean
\pi_0(n_\mu,n_5)&=&\pi^r(n_\mu,n_5)-\mu\tr P\tr[...\sigma_i U(n_\mu,n_5)...]\sigma^i
\ean
and solving for $\mu$ by plugging $\pi_0(n_\mu,n_5)$ into the hidden constraint (\ref{aq:cnstrP2})
\bean
\mu&=&\dfrac{\sum_{n_\mu,n_5}\tr\{\tr P\tr[...\sigma_iU(n_\mu,n_5)...]\sigma^i\pi^r(n_\mu,n_5)\}}{\sum_{n_\mu,n_5}\tr\{(\tr P\tr[...\sigma_i U(n_\mu,n_5)...]\sigma^i)^2\}}
\ean

\subsection{Rattle algorithm for the 5D orbifold gauge-Higgs model fixing $\langle\tr P\rangle$}\label{app:chmcorb}
\medskip
Given a constraint condition for the SU(2) link variables $U_5(n_\mu)$ in the fifth dimension
\bea
\dfrac{1}{2\Omega}\sum_{n_\mu,n_5}\prod_{n_5=0}^{N_5-1}[U_5(n_\mu,n_5)]\sigma_3\prod_{n_5=N_5-1}^{0}[U_5^\dagger(n_\mu,n_5)]\sigma_3=\Phi,\label{aq:orbcnst}
\eea
the constrained HMC (Rattle) algorithm can be formulated in the following way
\bsa
\pi_{n+1/2}&=&\pi_n-\dfrac{h}{2}\bigg(\dfrac{\partial S}{\partial U_n}-\dfrac{\lambda_n^{(1)}}{8\Omega}\tr[...\sigma_i U_n...-...U_n^\dagger\sigma_i...]\sigma^i\bigg)\\
U_{n+1}&=&e^{h\pi_{n+1/2}}U_n,\qquad U_{n+1}^\dagger=U_n^\dagger e^{-h\pi_{n+1/2}}\label{aq:orbUs}\\
0&=&\dfrac{1}{2\Omega}\sum_{n_\mu}\tr\prod_{n_5=0}^{N_5-1}[U_{n+1}(n_\mu,n_5)]\sigma_3\prod_{n_5=N_5-1}^{0}[U_{n+1}^\dagger(n_\mu,n_5)]\sigma_3-\Phi\label{ap:orbcnstr}\\
\pi_{n+1}&=&\pi_{n+1/2}-\dfrac{h}{2}\bigg(\dfrac{\partial S}{\partial U_{n+1}}-\dfrac{\lambda_n^{(2)}}{8\Omega}\tr[...\sigma_i U_{n+1}...-...U_{n+1}^\dagger\sigma_i...]\sigma^i\bigg)\label{aq:opin2}\\
0&=&\dfrac{1}{8\Omega}\sum_{n_\mu,n_5}\tr\{\tr[...\sigma_i U_{n+1}(n_\mu,n_5)...-...U_{n+1}^\dagger(n_\mu,n_5)\sigma_i...]\sigma^i\pi_{n+1}(n_\mu,n_5)\}\qquad \label{ap:orbcnstr2}
\esa

\noindent The first three lines determine $(\pi_{n+1/2},U_{n+1},\lambda_n^{(1)})$, whereas the remaining two give $(\pi_{n+1},\lambda_n^{(2)})$. Again, we truncate the exponentials in (\ref{aq:orbUs})
\bean
e^{h\pi_{n+1/2}}&=&1+h\pi_n-\dfrac{h^2}{2}\dfrac{\partial S}{\partial U_n}+\dfrac{h^2\lambda}{16\Omega}\tr[...\sigma_i U_n...-...U_n^\dagger\sigma_i...]\sigma^i+\dfrac{h^2}{2}\pi_n^2+\mathcal{O}(h^3)\\
e^{-h\pi_{n+1/2}}&=&1-h\pi_n+\dfrac{h^2}{2}\dfrac{\partial S}{\partial U_n}-\dfrac{h^2\lambda}{16\Omega}\tr[...\sigma_i U_n...-...U_n^\dagger\sigma_i...]\sigma^i+\dfrac{h^2}{2}\pi_n^2+\mathcal{O}(h^3)
\ean
and solve the constraint (\ref{ap:orbcnstr}) for the first Lagrange multipliers
\bea
\dfrac{\lambda_n^{(1)}}{8\Omega}&=&\bigg\{\sum_{n_\mu,n_5}\bigg(\tr[...\dfrac{\partial S}{\partial U_n(n_\mu,n_5)}U_n(n_\mu,n_5)...-...U_n^\dagger(n_\mu,n_5)\dfrac{\partial S}{\partial U_n(n_\mu,n_5)}...]\label{aq:orbl1}\\
&&\qquad-\tr[...\pi_n^2(n_\mu,n_5)U_n(n_\mu,n_5)...+...U_n^\dagger(n_\mu,n_5)\pi_n^2(n_\mu,n_5)...]\nn
&&\qquad-2\sum_{m_5}\tr\bigg[...\pi_n(n_\mu,n_5)U_n(n_\mu,n_5)...\pi_n(n_\mu,m_5>n_5)U_n(n_\mu,m_5>n_5)...\nn
&&\qquad\qquad\qquad-...\pi_n(n_\mu,n_5)U_n(n_\mu,n_5)...U_n^\dagger(n_\mu,m_5)\pi_n(n_\mu,m_5)...\nn
&&\qquad\qquad\qquad+...U_n^\dagger(n_\mu,m_5>n_5)\pi_n(n_\mu,m_5>n_5)...U_n^\dagger(n_\mu,n_5)\pi_n(n_\mu,n_5)...\bigg]\bigg)\bigg\}/\nn
&&\sum_{n_\mu,n_5}\tr[...\sigma_i U_n(n_\mu,n_5)...-...U_n^\dagger(n_\mu,n_5)\sigma_i...]\tr[...\sigma^iU_n(n_\mu,n_5)...-...U_n^\dagger(n_\mu,n_5)\sigma^i...]\nn
&&+\mathcal{O}(h^3)\non
\eea
We use a Secant method to get the Lagrange multiplier with machine precision. Plugging $\pi_{n+1}$ (\ref{aq:opin2}) into the hidden constraint (\ref{ap:orbcnstr2}) gives $\lambda_n^{(2)}$: 
\bea
\dfrac{\lambda_n^{(2)}}{8\Omega}&=&\sum_{n_\mu,n_5}\bigg(
\tr\{\tr[...\sigma_iU_n...-...U_n^\dagger\sigma_i...]\sigma^i\partial S/\partial U_{n+1}(n_\mu,n_5)\}\\
&&\qquad\quad-2\tr\{\tr[...\sigma_iU_n...-...U_n^\dagger\sigma_i...]\sigma^i\pi_{n+1/2}(n_\mu,n_5)\}/h\bigg)/\nn
&&\quad\sum_{n_\mu,n_5}\tr\{(\tr[...\sigma_i U_{n+1}...-...U_{n+1}^\dagger\sigma_i...]\sigma^i)^2\}\non
\eea

\noindent We initialize momenta $\pi_0$ from Gaussian-random distributed $\pi^r$ via orthogonal projection
\bean
\pi_0(n_\mu,n_5)&=&\pi^r(n_\mu,n_5)-\mu\tr[...\sigma_i U(n_\mu,n_5)...-...U(n_\mu,n_5)^\dagger\sigma_i...]\sigma^i\\
0&=&\dfrac{1}{8\Omega}\sum_{n_\mu,n_5}\tr\{\tr[...\sigma_i U(n_\mu,n_5)...-...U^\dagger(n_\mu,n_5)\sigma_i...]\sigma^i\pi_0(n_\mu,n_5)\}\\
\Rightarrow\mu&=&\dfrac{\sum_{n_\mu,n_5}\tr\{\tr[...\sigma_iU(n_\mu,n_5)...-...U(n_\mu,n_5)^\dagger\sigma_i...]\sigma^i\pi^r(n_\mu,n_5)\}}{\sum_{n_\mu,n_5}\tr\{(\tr[...\sigma_i U(n_\mu,n_5)...-...U(n_\mu,n_5)^\dagger\sigma_i...]\sigma^i)^2\}}
\ean

\newpage

\section{References}

\bibliographystyle{utphys}
\bibliography{paper2.bib}

\providecommand{\href}[2]{#2}\begingroup\raggedright\begin{thebibliography}{10}

\bibitem{Englert:1964et}
F.~Englert and R.~Brout, ``{Broken Symmetry and the Mass of Gauge Vector
  Mesons},''
\href{http://dx.doi.org/10.1103/PhysRevLett.13.321}{{\em Phys.Rev.Lett.}
  {\bfseries 13} (1964) 321--323}.

\bibitem{Higgs:1964ia}
P.~W. Higgs, ``{Broken symmetries, massless particles and gauge fields},''
\href{http://dx.doi.org/10.1016/0031-9163(64)91136-9}{{\em Phys.Lett.}
  {\bfseries 12} (1964) 132--133}.

\bibitem{ATLAS:2012gk}
{\bfseries ATLAS Collaboration} Collaboration, G.~Aad {\em et~al.},
  ``{Observation of a new particle in the search for the Standard Model Higgs
  boson with the ATLAS detector at the LHC},''
  \href{http://dx.doi.org/10.1016/j.physletb.2012.08.020}{{\em Phys.Lett.}
  {\bfseries B716} (2012) 1--29},
\href{http://arxiv.org/abs/1207.7214}{{\ttfamily arXiv:1207.7214 [hep-ex]}}.

\bibitem{CMS:2012gu}
{\bfseries CMS Collaboration} Collaboration, S.~Chatrchyan {\em et~al.},
  ``{Observation of a new boson at a mass of 125 GeV with the CMS experiment at
  the LHC},'' \href{http://dx.doi.org/10.1016/j.physletb.2012.08.021}{{\em
  Phys.Lett.} {\bfseries B716} (2012) 30--61},
\href{http://arxiv.org/abs/1207.7235}{{\ttfamily arXiv:1207.7235 [hep-ex]}}.

\bibitem{Manton:1979kb}
N.~Manton, ``{A New Six-Dimensional Approach to the Weinberg-Salam Model},''
\href{http://dx.doi.org/10.1016/0550-3213(79)90192-5}{{\em Nucl.Phys.}
  {\bfseries B158} (1979) 141}.

\bibitem{Fairlie:1979at}
D.~Fairlie, ``{Higgs' Fields and the Determination of the Weinberg Angle},''
\href{http://dx.doi.org/10.1016/0370-2693(79)90434-9}{{\em Phys.Lett.}
  {\bfseries B82} (1979) 97}.

\bibitem{Hosotani:1983vn}
Y.~Hosotani, ``{Dynamical Gauge Symmetry Breaking as the Casimir Effect},''
\href{http://dx.doi.org/10.1016/0370-2693(83)90841-9}{{\em Phys.Lett.}
  {\bfseries B129} (1983) 193}.

\bibitem{Irges:2004gy}
N.~Irges and F.~Knechtli, ``{Non-perturbative definition of five-dimensional
  gauge theories on the $\mathbb{R}^4\times S^1/\mathbb{Z}_2$ orbifold},''
  \href{http://dx.doi.org/10.1016/j.nuclphysb.2005.05.002}{{\em Nucl.Phys.}
  {\bfseries B719} (2005) 121--139},
\href{http://arxiv.org/abs/hep-lat/0411018}{{\ttfamily arXiv:hep-lat/0411018
  [hep-lat]}}.

\bibitem{Knechtli:2005dw}
F.~Knechtli, B.~Bunk, and N.~Irges, ``{Gauge theories on a five-dimensional
  orbifold},'' {\em PoS} {\bfseries LAT2005} (2006) 280,
\href{http://arxiv.org/abs/hep-lat/0509071}{{\ttfamily arXiv:hep-lat/0509071
  [hep-lat]}}.

\bibitem{Irges:2006zf}
N.~Irges and F.~Knechtli, ``{Non-perturbative mass spectrum of an
  extra-dimensional orbifold},''
\href{http://arxiv.org/abs/hep-lat/0604006}{{\ttfamily arXiv:hep-lat/0604006
  [hep-lat]}}.

\bibitem{Irges:2006hg}
N.~Irges and F.~Knechtli, ``{Lattice gauge theory approach to spontaneous
  symmetry breaking from an extra dimension},''
  \href{http://dx.doi.org/10.1016/j.nuclphysb.2007.01.023}{{\em Nucl.Phys.}
  {\bfseries B775} (2007) 283--311},
\href{http://arxiv.org/abs/hep-lat/0609045}{{\ttfamily arXiv:hep-lat/0609045
  [hep-lat]}}.

\bibitem{Elitzur:1979uv}
S.~Elitzur, R.~B. Pearson, and J.~Shigemitsu, ``{The Phase Structure of
  Discrete Abelian Spin and Gauge Systems},''
\href{http://dx.doi.org/10.1103/PhysRevD.19.3698}{{\em Phys. Rev.} {\bfseries
  D19} (1979) 3698}.

\bibitem{Ishiyama:2009bk}
K.~Ishiyama, M.~Murata, H.~So, and K.~Takenaga, ``{Symmetry and Z (2)
  Orbifolding Approach in Five-dimensional Lattice Gauge Theory},''
  \href{http://dx.doi.org/10.1143/PTP.123.257}{{\em Prog. Theor. Phys.}
  {\bfseries 123} (2010) 257--269},
\href{http://arxiv.org/abs/0911.4555}{{\ttfamily arXiv:0911.4555 [hep-lat]}}.

\bibitem{Irges:2013rya}
N.~Irges and F.~Knechtli, ``{Non-perturbative Gauge-Higgs Unification:
  Symmetries and Order Parameters},''
  \href{http://dx.doi.org/10.1007/JHEP06(2014)070}{{\em JHEP} {\bfseries 1406}
  (2014) 070},
\href{http://arxiv.org/abs/1312.3142}{{\ttfamily arXiv:1312.3142 [hep-lat]}}.

\bibitem{Irges:2012ih}
N.~Irges, F.~Knechtli, and K.~Yoneyama, ``{Mean-Field Gauge Interactions in
  Five Dimensions II. The Orbifold},''
  \href{http://dx.doi.org/10.1016/j.nuclphysb.2012.08.011}{{\em Nucl.Phys.}
  {\bfseries B865} (2012) 541--567},
\href{http://arxiv.org/abs/1206.4907}{{\ttfamily arXiv:1206.4907 [hep-lat]}}.

\bibitem{Irges:2012mp}
N.~Irges, F.~Knechtli, and K.~Yoneyama, ``{Higgs mechanism near the 5d bulk
  phase transition},''
  \href{http://dx.doi.org/10.1016/j.physletb.2013.04.032}{{\em Phys.Lett.}
  {\bfseries B722} (2013) 378--383},
\href{http://arxiv.org/abs/1212.5514}{{\ttfamily arXiv:1212.5514}}.

\bibitem{Alberti:2015pha}
M.~Alberti, N.~Irges, F.~Knechtli, and G.~Moir, ``{Five-Dimensional Gauge-Higgs
  Unification: A Standard Model-Like Spectrum},''
  \href{http://dx.doi.org/10.1007/JHEP09(2015)159}{{\em JHEP} {\bfseries 09}
  (2015) 159},
\href{http://arxiv.org/abs/1506.06035}{{\ttfamily arXiv:1506.06035 [hep-lat]}}.

\bibitem{Alberti:2016wff}
M.~Alberti, N.~Irges, F.~Knechtli, and G.~Moir, ``{Lines of Constant Physics in
  a Five-Dimensional Gauge-Higgs Unification Scenario},'' {\em PoS} {\bfseries
  LATTICE2016} (2016) 215,
\href{http://arxiv.org/abs/1609.07004}{{\ttfamily arXiv:1609.07004 [hep-lat]}}.

\bibitem{Fukuda:1974ey}
R.~Fukuda and E.~Kyriakopoulos, ``{Derivation of the Effective Potential},''
\href{http://dx.doi.org/10.1016/0550-3213(75)90014-0}{{\em Nucl. Phys.}
  {\bfseries B85} (1975) 354--364}.

\bibitem{ORaifeartaigh:1986axd}
L.~O'Raifeartaigh, A.~Wipf, and H.~Yoneyama, ``{The Constraint Effective
  Potential},''
\href{http://dx.doi.org/10.1016/S0550-3213(86)80031-1}{{\em Nucl. Phys.}
  {\bfseries B271} (1986) 653--680}.

\bibitem{Kuti:1987bs}
J.~Kuti and Y.~Shen, ``{Supercomputing the Effective Action},''
\href{http://dx.doi.org/10.1103/PhysRevLett.60.85}{{\em Phys. Rev. Lett.}
  {\bfseries 60} (1988) 85}.

\bibitem{Duane:1987de}
S.~Duane, A.~D. Kennedy, B.~J. Pendleton, and D.~Roweth, ``{Hybrid Monte
  Carlo},''
\href{http://dx.doi.org/10.1016/0370-2693(87)91197-X}{{\em Phys. Lett.}
  {\bfseries B195} (1987) 216--222}.

\bibitem{Irges:2017ztc}
N.~Irges and F.~Koutroulis, ``{Renormalization of the Abelian-Higgs model in
  the R $\xi$ and Unitary gauges and the physicality of its scalar
  potential},'' \href{http://dx.doi.org/10.1016/j.nuclphysb.2017.09.009}{{\em
  Nucl. Phys.} {\bfseries B924} (2017) 178--278},
\href{http://arxiv.org/abs/1703.10369}{{\ttfamily arXiv:1703.10369 [hep-ph]}}.

\bibitem{hairer:2002gni}
E.~Hairer, C.~Lubich, and G.~Wanner, {\em Geometric Numerical Integration.
  Structure-Preserving Algorithms for Ordinary Differential Equations}.
\newblock Springer, Berlin, 2nd ed.~ed., 2006.
\newblock \url{https://www.springer.com/de/book/9783540306634}.

\bibitem{hairer:2003gni}
E.~Hairer, C.~Lubich, and G.~Wanner, ``{Geometric numerical integration
  illustrated by the St\"ormer-Verlet method},''
  \href{http://dx.doi.org/10.1017/S0962492902000144}{{\em Acta Numerica}
  {\bfseries 12} (2003) 399--450}.
  \url{http://www.math.kit.edu/ianm3/lehre/geonumint2009s/media/gni_by_stoermer-verlet.pdf}.

\bibitem{Montvay:1994cy}
I.~Montvay and G.~Munster, {\em {Quantum fields on a lattice}}.
\newblock Cambridge University Press, 1997.
\newblock
\url{http://www.cambridge.org/uk/catalogue/catalogue.asp?isbn=0521404320}.
\newblock

\bibitem{Fodor:2007fn}
Z.~Fodor, K.~Holland, J.~Kuti, D.~Nogradi, and C.~Schroeder, ``{New Higgs
  physics from the lattice},'' {\em PoS} {\bfseries LATTICE2007} (2007) 056,
\href{http://arxiv.org/abs/0710.3151}{{\ttfamily arXiv:0710.3151 [hep-lat]}}.

\bibitem{Hollwieser:2018nrh}
{R. H\"ollwieser and F. Knechtli}, ``{Constraint HMC algorithms for gauge-Higgs
  models},'' {\em PoS} {\bfseries LATTICE2018} (2018) 052,
\href{http://arxiv.org/abs/1812.02045}{{\ttfamily arXiv:1812.02045 [hep-lat]}}.

\bibitem{Evertz:1986ur}
H.~G. Evertz, K.~Jansen, J.~Jersak, C.~B. Lang, and T.~Neuhaus, ``{Photon and
  Bosonium Masses in Scalar Lattice {QED}},''
\href{http://dx.doi.org/10.1016/0550-3213(87)90356-7}{{\em Nucl. Phys.}
  {\bfseries B285} (1987) 590--605}.

\bibitem{Ejiri:2000fc}
S.~Ejiri, J.~Kubo, and M.~Murata, ``{A Study on the nonperturbative existence
  of Yang-Mills theories with large extra dimensions},''
  \href{http://dx.doi.org/10.1103/PhysRevD.62.105025}{{\em Phys.Rev.}
  {\bfseries D62} (2000) 105025},
\href{http://arxiv.org/abs/hep-ph/0006217}{{\ttfamily arXiv:hep-ph/0006217
  [hep-ph]}}.

\bibitem{Knechtli:2011gq}
F.~Knechtli, M.~Luz, and A.~Rago, ``{On the phase structure of five-dimensional
  SU(2) gauge theories with anisotropic couplings},''
  \href{http://dx.doi.org/10.1016/j.nuclphysb.2011.11.001}{{\em Nucl.Phys.}
  {\bfseries B856} (2012) 74--94},
\href{http://arxiv.org/abs/1110.4210}{{\ttfamily arXiv:1110.4210 [hep-lat]}}.

\bibitem{Knechtli:2016pph}
F.~Knechtli and E.~Rinaldi, ``{Extra-dimensional models on the lattice},''
  \href{http://dx.doi.org/10.1142/S0217751X16430028}{{\em Int. J. Mod. Phys.}
  {\bfseries A31} no.~22, (2016) 1643002},
\href{http://arxiv.org/abs/1605.04341}{{\ttfamily arXiv:1605.04341 [hep-lat]}}.

\bibitem{deForcrand:2010be}
P.~de~Forcrand, A.~Kurkela, and M.~Panero, ``{The phase diagram of Yang-Mills
  theory with a compact extra dimension},''
  \href{http://dx.doi.org/10.1007/JHEP06(2010)050}{{\em JHEP} {\bfseries 1006}
  (2010) 050},
\href{http://arxiv.org/abs/1003.4643}{{\ttfamily arXiv:1003.4643 [hep-lat]}}.

\bibitem{Bergner:2015rza}
G.~Bergner, J.~Langelage, and O.~Philipsen, ``{Numerical corrections to the
  strong coupling effective Polyakov-line action for finite T Yang-Mills
  theory},'' \href{http://dx.doi.org/10.1007/JHEP11(2015)010}{{\em JHEP}
  {\bfseries 11} (2015) 010},
\href{http://arxiv.org/abs/1505.01021}{{\ttfamily arXiv:1505.01021 [hep-lat]}}.

\bibitem{Greensite:2017qfl}
{J. Greensite, and R. H\"ollwieser}, ``{Finite-density transition line for QCD
  with 695 MeV dynamical fermions},''
  \href{http://dx.doi.org/10.1103/PhysRevD.97.114504}{{\em Phys. Rev.}
  {\bfseries D97} no.~11, (2018) 114504},
\href{http://arxiv.org/abs/1708.08031}{{\ttfamily arXiv:1708.08031 [hep-lat]}}.

\bibitem{Hollwieser:2016hne}
{J. Greensite, and R. H\"ollwieser}, ``{Relative weights approach to SU(3)
  gauge theories with dynamical fermions at finite density},''
  \href{http://dx.doi.org/10.1103/PhysRevD.94.014504}{{\em Phys. Rev.}
  {\bfseries D94} no.~1, (2016) 014504},
\href{http://arxiv.org/abs/1603.09654}{{\ttfamily arXiv:1603.09654 [hep-lat]}}.

\end{thebibliography}\endgroup

\end{appendices}

\end{document}